\documentclass[aps,prl,twocolumn,
groupedaddress,showpacs,amsmath,amssymb,longbibliography, amsfonts,10pt]{revtex4-2}

\usepackage{amsmath}
\usepackage{amsfonts}
\usepackage{amssymb}
\usepackage{graphicx}
\usepackage{color}
\usepackage{xcolor}
\usepackage{bbold}
\usepackage{appendix}
\usepackage[citecolor=blue,urlcolor=blue]{hyperref}
\hypersetup{colorlinks}
\usepackage[caption=false]{subfig}
\usepackage{bookmark}
\usepackage{makecell}
\usepackage{mathtools}
\usepackage{comment}
\usepackage{tikz}
\usetikzlibrary{arrows.meta}
\usepackage{array}
\usepackage{multirow}
\usepackage[nolist,nohyperlinks]{acronym}
\usepackage[export]{adjustbox}
\usepackage[normalem]{ulem}
\usepackage{svg}
\usepackage{placeins}
\graphicspath{{./figures/}}
\newacro{RG}[RG]{renormalization group}
\newacro{fdr}[FDR]{fluctuation-dissipation relations}
\newacro{EP}[EP]{exceptional point}

\newacro{CEP}[CEP]{critical exceptional point}
\newacro{frg}[FRG]{functional RG}
\newacro{DSE}[DSE]{Dyson-Schwinger equation}
\newacro{MSRJD}[MSRJD]{Martin-Siggia-Rose-Janssen-de Dominicis}
\newacro{BKT}[BKT]{Berezinskii–Kosterlitz–Thouless}
\newacro{KPZ}[KPZ]{Kardar–Parisi–Zhang}
\newacro{RWA}[RWA]{rotating wave approximation}
\newacro{Nlsm}[NL$\sigma$M]{non-linear $\sigma$ model}
\newacro{SM}[SM]{Supplemental Materials}
\usepackage{lipsum}

\definecolor{rotatingColor}{rgb}{0.741176, 0.811765, 1}
\definecolor{xyColor}{rgb}{0.611765, 0.905882, 0.658824}
\definecolor{paraColor}{rgb}{0.952941, 0.952941, 0.913725}
\definecolor{cepColor}{rgb}{0.858824, 0, 0}

\begin{document}
\title{Nonequilibrium criticality at the onset of time-crystalline order}
\author{Romain Daviet} 
\author{Carl Philipp Zelle}
\author{Achim Rosch}
\author{Sebastian Diehl}

\affiliation{Institut f\"ur Theoretische Physik, Universit\"at zu K\"oln, 50937 Cologne, Germany}

\begin{abstract}
We explore the phase transitions at the onset of time-crystalline order in $O(N)$ models driven out-of-equilibrium. The spontaneous breaking of time-translation symmetry and its Goldstone mode are captured by an effective description with $O(N)\times SO(2)$ symmetry, where the emergent external $SO(2)$ results from a transmutation of the internal symmetry of time translations. 
Using the renormalization group and the $\epsilon=4-d$ expansion in a leading two-loop analysis, we identify a new nonequilibrium universality class. Strikingly, it controls the long-distance physics no matter how small the microscopic breaking of equilibrium conditions is.  The $O(N=2)\times SO(2)$ symmetry group is realized for magnon condensation in pumped yttrium iron garnet  films and in exciton-polariton systems with a polarization degree of freedom.
\end{abstract}

\date{\today}

\maketitle
\paragraph{Introduction -- }

 Temporal pattern formation, associated with instabilities at a finite frequency such as waves, fronts, or oscillatory behaviors in classical systems~\cite{Cross1993}, represents a class of genuine nonequilibrium phenomena. In a many-body setting, these patterns can become robust against fluctuations and synchronize over large time and spacescales. The ensuing long-range order is then associated with the spontaneous breaking of time-translation symmetry. When the system retains a periodic motion, the order parameter describes a limit cycle. In other words, it corresponds to a time crystal~\cite{Shapere2012} for which a quantum version was proposed~\cite{Wilczek2012}, and rapidly shown to exist only out of equilibrium~\cite{Bruno2013,Watanabe2015}.
Because they provide a generic way of breaking equilibrium conditions at microscopic scales, driven open systems turned out to be a prolific arena to study and realize finite frequency instabilities~\cite{Scarlatella2019}, limit cycles~\cite{Piazza2015,Dogra2019,Buca2022,Buca2019} and discrete time crystals~\cite{Gong2018,Kessler2021}, characterized by a subharmonic response at a multiple of the period of an external periodic drive~\cite{Khemani2016,Else2016,Choi2017,Zhang2017,Yao2020,Zaletel2023}. These systems also opened up the search for dissipative continuous time crystals~\cite{Iemini2018,Buca2019,Kessler2019}, which were experimentally realized~\cite{Kongkhambut2022}. At the same time, active matter systems~\cite{Marchetti2013} have become platforms to describe many-body temporal instabilities, as they share the generic breaking of equilibrium conditions with driven systems.

The phase transition into a time crystal, concomitant with the  spontaneous breaking of time-translation symmetry, must be expected to display phenomena that fall beyond equilibrium classifications, by the very nature of the adjacent nonequilibrium phase. In this Letter, we study these for paradigmatic $O(N)$ symmetric models in the absence of conservation laws. 
Once suitably brought out of equilibrium, they develop instabilities toward time-crystalline long-range orders, which emerge in nonreciprocal active matter~\cite{Fruchart2021,zelle23} as well as in more generic driven quantum and solid-state platforms~\cite{zelle23, Hanai2019}.
Time translation symmetry breaking occurs in two distinct scenarios, one generalizing Van der Pol oscillations and another one describing an order parameter rotating in the $O(N)$ manifold. A key insight of our approach lies in a fruitful mapping to an effective theory with an $O(N) \times SO(2)$ symmetry, where the emergent $SO(2)$ arises from time-translations combined with the finite frequency scale set by the periodicity of the limit cycles~\cite{Cross1993}. This setup provides us with a complete understanding of the Goldstone modes and their dynamics, including the one associated with broken time-translation symmetry~\cite{hayata2018}. The interplay of the latter with those stemming from the broken $O(N)$ symmetry~\cite{zelle23} materializes differently in the Van der Pol oscillating and the rotating phases. By performing a perturbative \ac{RG} analysis of the effective theory, we obtain the full phase diagram beyond mean field by including fluctuation effects. It leads us to our main results: The onset of the Van der Pol oscillations is governed by fluctuation-induced first-order transitions for $N>1$. Conversely, the transition to the rotating phase resides in a fundamentally new nonequilibrium universality class, not smoothly connected to any known equilibrium class. We also exhibit realization of the symmetry group $O(N=2)\times SO(2)$ in exciton-polariton systems with a polarization degree of freedom and in magnon
condensation in pumped yttrium iron garnet (YIG) films.

\paragraph{Time-crystalline orders--}
 Our starting point is the following Langevin equation describing the dynamics of an $O(N)$ symmetric order parameter $\boldsymbol{\phi}(\boldsymbol{x},t)\in\mathbb{R}^N$ in $d$ spatial dimensions,
\begin{align}
\label{eq:EoMbegin}
\begin{split}
    &\partial_t^2\boldsymbol{\phi}+(2\gamma+u\rho-Z_1\nabla^2)\partial_t\boldsymbol{\phi}+\frac{u'}{2}\partial_t\rho\,\boldsymbol{\phi}\\ &+(r+\lambda\rho-Z_2\nabla^2)\boldsymbol{\phi}+\boldsymbol{\xi}=0,
    \end{split}
\end{align}
where $\rho=\boldsymbol{\phi} \cdot \boldsymbol{\phi}$ and $\boldsymbol{\xi}(\boldsymbol{x},t)$ is a Gaussian white noise with zero mean and variance $\langle \xi_i(\boldsymbol{x},t)\xi_j(\boldsymbol{x}',t')\rangle= 2D\delta_{i,j}\delta(\boldsymbol{x}-\boldsymbol{x}')$. This constitutes a generalization of the Van der Pol oscillator~\cite{vanderPol1920}, often advocated to describe both classical and quantum limit cycles ~\cite{Walter2014,Walter2014a,Lee2013,Dutta2019,BenArosh2021}, with an $N$-component field, an extensive number of spatial degrees of freedom, and noise. We consider the most relevant operators compatible with the $O(N)$ symmetry -- they are generated upon coarse-graining. In addition to the potential $r+\lambda\rho$, there are nonlinear dampings $u$ and $u'$, a spatially dependent damping $Z_1$, and diffusion $Z_2$. The noise encodes random fast fluctuations coming from the environment (e.g., drive, bath). While these fluctuations are typically counteracted by a finite gap $r$ or a finite damping $\gamma$, they strongly affect the dynamics near a phase transition. 

For negative values of $r$ in Eq.~\eqref{eq:EoMbegin}, the usual static equilibrium $O(N)$ order parameter $\boldsymbol{\phi}_S= \langle \boldsymbol{\phi}(\boldsymbol{x},t)\rangle$ builds up, $|\boldsymbol{\phi}_S|=\sqrt{-r/\lambda}$. The $O(N)$ symmetry is spontaneously broken to $O(N-1)$ via the equilibrium model A transition~\cite{Hohenberg1977}. 
Here, we follow a different route to generate order, and tune the damping $\gamma$ to negative values while keeping $r$ strictly positive, $r>0$. In a driven setting, such an antidamping, $\gamma < 0$, lends itself to the intuition for a nonequilibrium situation, where losses are superseded by pumping.

\begin{figure}
    \centering
    \includegraphics{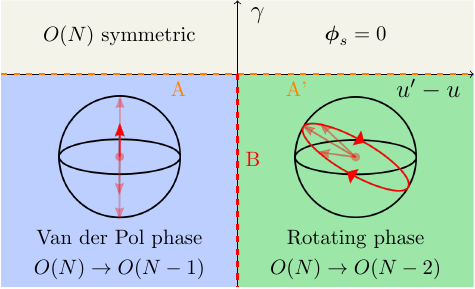}
    \caption{Mean-field phase diagram for $r>0$. Tuning the damping to negative values stabilizes two time-crystal phases, depending on the sign of $u'-u$. The symmetric and limit-cycle phases are separated by the transition lines A and A', which are second-order transitions at mean field. The line B separates the ordered phases and is a first-order transition, since the order parameter jumps from one configuration to the other. It is well described within mean-field approximation, unlike the other two transitions. Analysis of the latter is the main focus of this Letter.}
    \label{fig:phaseDiag}
\end{figure}

For $\gamma<0$, there are two time-periodic solutions of the noise-averaged version of Eq.~\eqref{eq:EoMbegin}, stabilized by the additional nonlinearities $u$ and $u'$~\cite{zelle23}.
The first one exists only for $N>1$, and is a rotating phase where the order parameter traces out a circle on the $N$-sphere whose orbit is spontaneously chosen, see Fig.~\ref{fig:phaseDiag}, $\boldsymbol{\phi}_S(t)= \sqrt{-2 \gamma/u} (\cos(\omega_0 t),-\sin(\omega_0 t),0,\dots,0)^T$ with $\omega_0= \sqrt{r-\frac{2\gamma\lambda}{u}}$. The second phase is characterized by an oscillation of the order parameter along one direction, and exists for any $N$. We refer to it as the Van der Pol phase since the order parameter reads as $\boldsymbol{\phi}_S(t)= \phi_{0}(t) (1,0,\dots,0)^T$,
where $\phi_0(t)$ is the limit cycle of the Van der Pol equation~\footnote{This is formally true only when $\lambda=0$, but this term does not impact the mean-field picture.}. At lowest order in $|\gamma| \ll \sqrt{r}$~\cite{Verhulst1996}, the solution is $\phi_{0}(t) \approx 2 \sqrt{\frac{-2\gamma}{(u+u')}}\cos(\omega_0 t) $, with $\omega_0 = \sqrt{r}$. 
For $\gamma<0$, only one of the phases is stable for a given set of parameters. A stability analysis reveals that the rotating phase is stable for sufficiently large values of $u'$~\cite{zelle23}. For $\gamma \to 0$ and $r>0$, the rotating phase is stable for $u'>u$. This leads to the phase diagram in Fig.~\ref{fig:phaseDiag}.

\paragraph{Symmetry breaking patterns and Goldstone modes --} The Van der Pol phase spontaneously breaks the $O(N)$ symmetry to $O(N-1)$, leading to $N-1$ Goldstone modes, while the rotating phase breaks $O(N)$ to $O(N-2)$, leading to $2N-3$ Goldstone modes~\cite{zelle23}.
In both phases, time-translation symmetry is spontaneously broken to discrete time-translation symmetry. Indeed, the solutions are periodic, and a time translation $t \to t+2\pi n/\omega_0 ,\ n\in \mathbb{Z}$ leaves them invariant. They are, therefore, nonequilibrium time-crystal phases. This is striking for $N=1$, where a continuous symmetry is broken rather than the discrete $O(N=1) \simeq\mathbb{Z}_2$ symmetry~\footnote{A $O(1)=\mathbb{Z_2}$ transformation $\boldsymbol{\phi}_S \to -\boldsymbol{\phi}_S$ can always be compensated by a time translation by half the period. The $\mathbb{Z}_2$ symmetry is therefore unbroken.}.
 
Time translation is a continuous symmetry, and a Goldstone mode arises from its spontaneous breaking. We show in the \ac{SM}~\footnote{See Supplemental Material at [URL], which includes Refs.~\cite{ZinnJustin,Kleinert2001,Dupuis2022,Bausch1976,Andreanov2006,Aron2010,Gyoergyi1992,Aron2016,Ohadi2015,Pelissetto2002,Martin1973,Janssen1976,Dominicis1976,Kawamura1988,Ohmi1998,Ho1998,Debelhoir2016,Rezende2009}, for a detailed discussion of the field theoretical treatment of the Langevin equations discussed in the text.} that the Goldstone theorem gives an additional Goldstone mode in the Van der Pol phase, which therefore has a $N$ Goldstone modes. In the rotating phase, a distinct scenario occurs: a rotation along the orbit is fully equivalent to a time translation. The rotating phase has a remaining $SO(2)$ symmetry that acts via $\boldsymbol{\phi} \to R(\alpha) \boldsymbol{\phi}, \quad t \to t-\frac{\alpha}{\omega_0}$, with $\alpha \in \mathbb{R}, R \in SO(2)$, and no independent Goldstone mode arises: The Goldstone mode of time-translation symmetry is redundant and reflects the activation of a rotational Goldstone mode.

\paragraph{Gaussian fluctuations phase transitions --}

The transitions from the disordered regime into the limit-cycle phases are reached as the damping $\gamma$ goes to zero, see Fig.~\ref{fig:phaseDiag}. The amplitudes of the order parameters are continuous at the transitions, but the oscillations immediately start with a frequency $\omega_0 \sim \sqrt{r}$, which acts as a finite and fast timescale close to the transition, where $|\gamma| \ll \sqrt{r}$. We first neglect the effects of the nonlinearities. It is a valid assumption for $d>d_c$, the upper critical dimension, which we determine to be equal to four below.

The transition is characterized by the retarded response to an external field $h$, $\chi^R(\boldsymbol{x},t)=\frac{\delta \langle \phi_i(\boldsymbol{x},t)\rangle}{\delta h_i(0,0)}$ and the correlation function
\begin{align}
\label{eq:resp_corr_first}
    \langle \phi_i(\boldsymbol{q},t)\phi_i(-\boldsymbol{q},0)\rangle &\sim D\frac{e^{-(\frac{Z_1}{2}q^2+\gamma)|t|}}{\frac{Z_1}{2}q^2+ \gamma} \cos (\sqrt{r}t+\frac{Z_2}{2\sqrt{r}}q^2 t) ,
\end{align}
with $q=|\mathbf{q}|$.
The correlation function displays an algebraic divergence as $q\rightarrow 0$, characteristic of a second-order phase transition, at $\gamma=0$. The diverging correlation length, $\xi_c\sim \gamma^{-1/2}/Z_1$, leads to the mean-field critical exponent $\nu=1/2$. There are, however, oscillations with frequency $\omega_0 \sim \sqrt{r}$, and the divergence occurs at finite frequencies $\omega\sim \pm \omega_0$.
This hinders a direct RG analysis of the transition~\footnote{We explicitly checked that the two frequencies spoil the generic structure of RG study based on series expansion around a given momentum and frequency. This leads to nonphysical divergences, technically reminiscent of what was found in~\cite{zelle23}.}, and we first need to distill the slow degrees of freedom, eliminating the fast scale $\omega_0$\cite{Cross1993}. To do this, we first pass to a system of first order differential equations by introducing $\boldsymbol{\Pi}=\partial_t \boldsymbol{\phi}/\omega_0$. We then define new $O(N)$ variables $(\boldsymbol{\chi}_1,\boldsymbol{\chi}_2)$ via a rotation by an angle $\omega_0 t$ of the original variables $(\boldsymbol{\phi},\boldsymbol{\Pi})^T=R(\omega_0 t)(\boldsymbol{\chi}_1,\boldsymbol{\chi}_2)^T$, or explicitly   
\begin{equation}
\begin{split}
    \label{eq:defphi12}
    \boldsymbol{\phi}(\boldsymbol{x},t)&=\boldsymbol{\chi}_1(\boldsymbol{x},t)\cos(\omega_0t)+\boldsymbol{\chi}_2(\boldsymbol{x},t)\sin(\omega_0t),\\
    \boldsymbol{\Pi}(\boldsymbol{x},t)&=-\boldsymbol{\chi}_1(\boldsymbol{x},t)\sin(\omega_0t)+\boldsymbol{\chi}_2(\boldsymbol{x},t)\cos(\omega_0t),
    \end{split}
\end{equation}
 where $\boldsymbol{\chi}_{1,2}$ vary slowly compared to the scale $\omega_0$, with their main fluctuations concentrated around zero momentum and frequency.
 Subsequently, since the critical behavior occurs on timescales $t_c \sim \gamma^{-1}\gg \omega_0^{-1}$, we treat the fast oscillating terms that appear by averaging over time in a \ac{RWA}~\cite{Verhulst1996}. The resulting near-critical Langevin equations for $\boldsymbol{\chi}_1$ and $\boldsymbol{\chi}_2$ are then no longer explicitly time dependent.
 
\paragraph{Effective $O(N)\times SO(2)$ model --}
Remarkably, the dynamics obtained after the \ac{RWA} display an emergent $SO(2)$ symmetry on top of the $O(N)$ symmetry,  
\begin{equation}
\label{eq:defRotSO2}
    \left(\begin{array}{cc}
         \boldsymbol{\chi}_1  \\
         \boldsymbol{\chi}_2 
    \end{array}\right) \to 
        R(\alpha)\left(\begin{array}{cc}
         \boldsymbol{\chi}_1  \\
         \boldsymbol{\chi}_2 
    \end{array}\right), \quad \text{where }  R(\alpha) \in SO(2)  
\end{equation}
is a rotation by an angle $\alpha$. It roots in the fact that a constant arbitrary shift $\omega_0  t \to\omega_0  t + \alpha$ in the rotation~\eqref{eq:defphi12} defining $\boldsymbol{\chi}_1$ and $\boldsymbol{\chi}_2$ does not change the final equations of motion. 
We observe a transmutation of the external time-translation symmetry to an internal $SO(2)$~\cite{Cross1993} acting in the space of fields $\boldsymbol{\chi}_1,\boldsymbol{\chi}_2$. This is rationalized by the fact that time translations can be identified modulo $2\pi/\omega_0$ upon approaching a limit cycle: The group of time translations is isomorphic to $\mathbb{R}$, while the discrete group that leaves the limit cycles invariant has a $\mathbb{Z}$ structure. It follows that the relevant subpart of time translations that gets broken is $\mathbb{R}/\mathbb{Z} \simeq SO(2)$.

The equations of motion are
\begin{align}
 \label{eq:EoM_12}
    &\partial_t \boldsymbol{\chi}_a + \frac{\delta H_{d}}{\delta \boldsymbol{\chi}_a}+ \epsilon_{ab}\frac{\delta H_{c}}{\delta \boldsymbol{\chi}_b}+\boldsymbol{\xi}_a=0, \quad (a,b) \in \{1,2\}\\
    &H_{l}=\int d^d \boldsymbol{x}\frac{Z_{l}}{2}\left[\left(\boldsymbol{\nabla} \boldsymbol{\chi}_1\right)^2+\left(\boldsymbol{\nabla} \boldsymbol{\chi}_2\right)^2\right]+\frac{\gamma_l}{2} \rho+\frac{g_l}{8} \rho^2+\frac{\kappa_l}{2} \tau,\nonumber
\end{align}
 with $l \in \{c,d\}$, $\boldsymbol\xi_a$ two independent noises, and the two $O(N)\times SO(2)$ invariants:
$\rho=\boldsymbol{\chi}_1^2+\boldsymbol{\chi}_2^2$ and 
$\tau=\frac{1}{4}\left(\boldsymbol{\chi}_1^2-\boldsymbol{\chi}_2^2\right)^2+\left(\boldsymbol{\chi}_1 \cdot \boldsymbol{\chi}_2\right)^2$.
From Eq.~\eqref{eq:EoMbegin}, one obtains $\gamma_c=0$, $\gamma_d=\gamma$, $Z_c=Z_2/2 \omega_0$, $Z_d=Z_1/2$, $g_d= u/2$, $\kappa_d=(u'-u)/4$, $g_{c}=\lambda/2 \omega_0$ and $\kappa_{c}=\lambda/4\omega_0$. The fast scale is implicit, confirming that we have isolated the slow degrees of freedom.

We note here a parallel to frustrated magnets and helimagnets~\cite{Kawamura1998} at equilibrium, for which the order parameters take a similar form as~\eqref{eq:defphi12}, with space playing the role of time. These systems have an emergent $O(N)\times O(2)$ symmetry, and Hamiltonian $H_d$. This suggests that the difference with an equilibrium situation is linked to $H_{c}$ in Eq.~\eqref{eq:EoM_12}, which describes coherent effects rather than dissipation. Indeed, it is this term that explicitly breaks $O(2)$ to $SO(2)$. Its presence is closely, but not exactly, related to the breaking of equilibrium conditions: The system is in thermal equilibrium if and only if $H_c$ is proportional to $H_d$~\cite{Sieberer2013,Taeuber2014a,Note3,Dominicis1975}. 

\begin{figure}
    \centering
 \includegraphics[width=7.75cm]{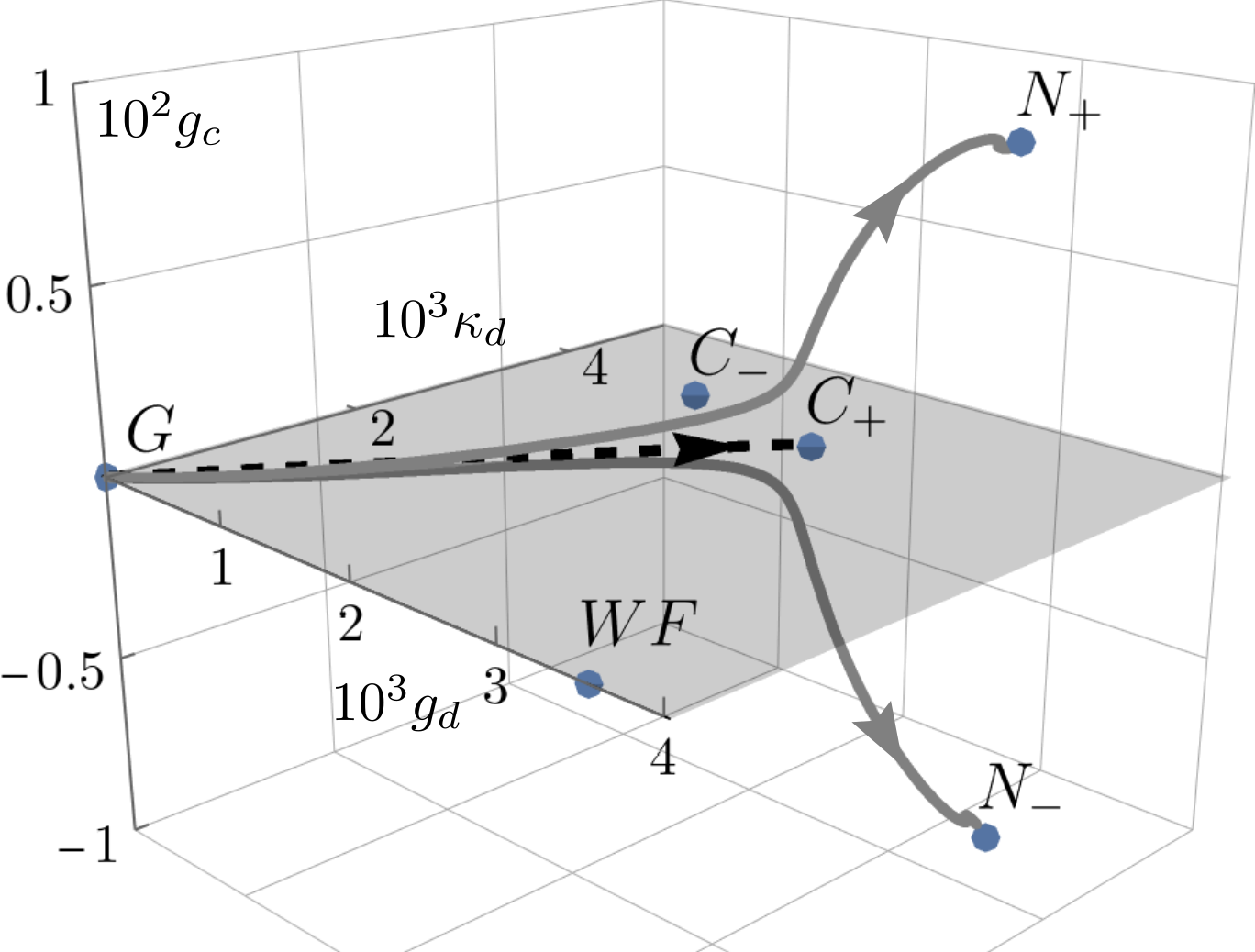}
   \caption{Flow diagram at criticality ($\gamma=0$) projected on the $( g_d, \kappa_d,g_c)$ manifold, for $N=24$ and $\epsilon=0.1$. Similar flows are obtained for all $N>1$. In addition to the Gaussian fixed point, there are three equilibrium fixed points~\cite{Kawamura1998,Calabrese2004,Delamotte2004}, labeled Wilson-Fisher (WF), chiral ($C_+$), and antichiral ($C_-$). They lie in the $g_c=0$ gray plane. $C_+$ is attractive at equilibrium, and the equilibrium black-dashed trajectory is attracted toward it. However, a small breaking of equilibrium grows at larger scale, and the gray solid trajectories go to the new nonequilibrium complex conjugated fixed points $N_{\pm}$.
   }
    \label{fig:FlowDiagMain}
\end{figure}

We can obtain an equivalent complex representation of Eq.~\eqref{eq:EoM_12}, which makes contact with the semiclassical dynamics of driven open quantum systems~\cite{Sieberer2016,sieberer2023universality}. In terms of the complex vector field $\boldsymbol{\psi} = \boldsymbol{\chi}_1+ i\boldsymbol{\chi}_2 \in \mathbb{C}^N$, and noise $\boldsymbol{\xi} \in \mathbb{C}^N$, Eq.~\eqref{eq:EoM_12} becomes
\begin{align}
\label{eq:EoM_complex}
    (i\partial_t -Z\nabla^2  + i\gamma)\boldsymbol{\psi} + \frac{g}{2} (\boldsymbol{\psi} \cdot \boldsymbol{\psi}^{*}) \boldsymbol{\psi} + \frac{\kappa}{2}  (\boldsymbol{\psi} \cdot \boldsymbol{\psi})\boldsymbol{\psi}^*+ \boldsymbol{\xi} =0,
\end{align}
 with $Z=i Z_d+ Z_c$, $g=i g_d+ g_c$ and $\kappa=i\kappa_d+ \kappa_c$. Eq.~\eqref{eq:EoM_complex} is a generalized noisy Gross-Pitaevskii equation, where the imaginary parts encode the effect of drive and dissipation on top of the coherent Hamiltonian dynamics. 
The $N=1$ case, with $U(1)\simeq SO(2)$ symmetry~\footnote{Again, the $O(1)= Z_2$ symmetry is redundant~\cite{Note2}.}, describes the dynamics of driven-dissipative Bose-Einstein condensates ~\cite{Sieberer2013,Sieberer2014,Taeuber2014a}, and those of collections of classical oscillators~\cite{Risler2004,Risler2005}.

 We therefore obtain a first nontrivial result. The transition to the Van der Pol phase for $N=1$ coincides with the transition in those systems: It has an emergent thermal equilibrium and falls into the $O(2)$ model A universality class, albeit with an additional exponent describing the fade out of coherent dynamics~\cite{Sieberer2013,Sieberer2014,Taeuber2014a}. 
 The $N>1$ case, however, differs notably as we will see. 
 
\paragraph{Phase diagram revisited --}

The mean-field analysis is fully recovered from the effective Langevin equation~\eqref{eq:EoM_12} or~\eqref{eq:EoM_complex}. The limit cycles are described by finite expectation values of $\boldsymbol{\chi}_1$ and $\boldsymbol{\chi}_2$, see Eq.~\eqref{eq:defphi12}. They are found by solving $\delta H_{d}/\delta{\boldsymbol{\phi}_a}=0$, while the nonzero remaining part coming from $H_{c}$ can always be canceled by a 
redefinition of $\omega_0$, see SM~\cite{Note3}. There are two ways of minimizing $H_{d}$ for $\gamma<0$, depending on the value of $\kappa_d$: If it is positive, the rotating phase is favored with $\tau=0$ i.e., $\boldsymbol{\chi}_1$,  $\boldsymbol{\chi}_2$ nonzero and orthogonal. When $\kappa_d<0$, the stable state has $\tau\neq 0$ i.e, $\boldsymbol{\chi}_1$ and $\boldsymbol{\chi}_2$ parallel, which corresponds to the Van der Pol phase.

The correlation functions agree as well,
\begin{equation}
    C(\boldsymbol{q},t) =\langle \psi_i \psi_i^* \rangle (\boldsymbol{q},t)\propto \frac{ D \ e^{-(Z_d q^2+\gamma- i Z_c q^2)|t|}}{Z_d q^2+\gamma}, 
\end{equation}
 and Eq.~\eqref{eq:resp_corr_first} is reproduced up to the, now implicit, fast scale. This allows us to make a scaling ansatz~\cite{Taeuber2014,Taeuber2014a,Note3},
 \begin{align}
     \label{eq:scalingAnsatz}
     \chi^R(\boldsymbol{q},t) &\sim q^{-2+\eta'+z}\tilde \chi^R(t q^z, i q^{\eta-\eta_c},q\gamma^{-\nu}),\\
     C(\boldsymbol{q},t) &\sim q^{-2+\eta}\tilde C(t q^z,i q^{\eta-\eta_c},q\gamma^{-\nu}), \label{eq:scalingAnsatz2}
 \end{align}
with critical exponents $\nu=1/2$, $z=2$, $\eta=\eta'=\eta_c=0$ at mean field. $\eta_c$ encodes the competition between coherent and dissipative effects~\cite{Sieberer2014}, while $\eta'\neq \eta$ entails a violation of fluctuation-dissipation relations, which relate responses and correlations in equilibrium. Below, we will compute the fluctuation corrections.
 
Besides recovering the number of Goldstone modes, we obtain their explicit dynamics and dispersion relations $\omega(q)$, see SM~\cite{Note3}. They are dissipative in the Van der Pol phase, $\omega(q)= -i Z_d q^2$. In the rotating phase, the Goldstone mode associated with time-translation breaking is also dissipative, while the others display both dissipative and coherent parts, $\omega(q)= -i Z_d q^2 \pm Z_c q^2$. These dispersions imply divergent fluctuations in $d\leq 2$, which destroy the ordered phases at long distances, even for $N=1$.
\begin{table}
\renewcommand{\arraystretch}{1.25}
\setlength{\tabcolsep}{6pt}
\begin{tabular}{|c|c|c|c|c|}
    \hline
   $N$ & $\nu^{-1}-2$ &$\eta=\eta_c$& $10^{3}(z-2)$ & $10^{2}\eta'$\\\hline
     $22$ &  $-0.942\epsilon$ &$-0.142 \epsilon^ 2$ &$5.5  \epsilon^ 2$  & $(0.030+ 1.8i)\epsilon^2$ \\\hline
     $3$  & $-1.27\epsilon$ & $-1.49 \epsilon^ 2$& $-1.7 \epsilon^ 2$&  $(-3.5+ 6.7i)\epsilon^2$ \\\hline
     $2$  &  $-0.853\epsilon$ & $-0.353 \epsilon^ 2$ &$ 7.2 \epsilon^ 2$ &  $(1.0+ 0.70 i)\epsilon^2$\\\hline
\end{tabular}
\caption{Critical exponents of the attractive fixed points $N_\pm$ controlling the transition to the rotating phase for different values of $N$ to lowest order in $\epsilon$. }

\label{tab:crit_exp}
\end{table}

\paragraph{Phase transition in $d<4$ -- } We now use dynamical \ac{RG} to treat the effects of interactions and fluctuations.
It allows us to identify nontrivial scaling regimes as well as fluctuation-induced first-order phase transitions, which are found whenever no fixed point can be reached within the RG flow~\cite{Amit2005}. 
We compute the flow equations in the $\epsilon=4-d$ perturbative expansion. The parameter $r_K=Z_c/Z_d$ enters the RG equations, and a two-loop analysis of the self-energies is needed to fully characterize the fixed points. 
The flow equations are derived in the \ac{SM}~\cite{Note3}. Interactions are irrelevant above $d_c=4$, but nontrivial fixed points emerge below it. A flow diagram is displayed in Fig.~\ref{fig:FlowDiagMain}. 

First, equilibrium fixed points, associated with the $O(N)\times O(2)$ symmetry~\cite{Kawamura1998,Calabrese2004,Delamotte2004}, are still solutions with $g_c=\kappa_c= r_K=0$. Now, if the initial conditions correspond to equilibrium, i.e., $g_c/g_d=\kappa_c/\kappa_d= r_K$, the system remains at equilibrium at all scales. 
However, one key finding is that the equilibrium fixed points are unstable against any, even infinitesimal, nonequilibrium perturbations. Indeed, these equilibrium fixed points acquire a relevant direction, associated with the microscopic breaking of equilibrium conditions, and none of them controls the transition. 
This highly unusual behavior at a second-order phase transition can be rationalized by the fact that we describe the onset of genuine nonequilibrium phases.

Instead, the flow is attracted toward a pair of new fixed points $N_{\pm}$, not present in the $O(N)\times O(2)$ case, that govern the transition. Since the couplings do not have a fixed ratio between imaginary and real parts e.g., $g_c/g_d \neq r_K$, thermal conditions are violated at the fixed points. They thus define a new nonequilibrium universality class, uniquely associated with the $O(N)\times SO(2)$ symmetry.
They exist for any value of $N$, in sharp contrast with the equilibrium ones~\cite{Kawamura1998,Calabrese2004,Delamotte2004}. They are complex conjugates, i.e., they describe mutually time-reversed coherent dynamics in~Eq.~\eqref{eq:EoM_complex}~\cite{Marino2016,Marino2016a,Young2020}.
We find that the attractive fixed points have $\kappa_d>0$ for $N>N_{c} \sim 1.6+ O(\epsilon)$ and $\kappa_d<0$ otherwise. It means that the transition to the Van der Pol phase is fluctuation-induced first-order for every $N>1$ (and described by the equilibrium $O(2)$ transition for $N=1$), while the transition to the rotating phase is second-order and described by the uncovered universality class for all $N$, at least close to four dimensions.
The critical exponents to leading order in $\epsilon$ of this universality class, distinguishing it from any known class, are given in Tab.~\ref{tab:crit_exp}. Its nonequilibrium nature is reflected in $\eta'\neq\eta$.

\paragraph{Conclusion --} 
Our results can be applied to a wide range of physical situations.
In addition to the $O(N)$ symmetric systems described by Eq.~\eqref{eq:EoMbegin}, one can start directly from Eq.~\eqref{eq:EoM_complex}. For $N=2$, one needs two driven-dissipative complex bosonic degrees of freedom $\psi_\pm$ connected by an exchange symmetry $\psi_+\leftrightarrow \psi_-$. Together, this forms the group $O(2)\times SO(2)$, see~\ac{SM}~\cite{Note3}. Remarkably, this symmetry is obtained in existing platforms, known to exhibit driven Bose-Einstein condensation described by noisy Gross-Pitaevskii equations. A point in case are exciton-polaritons, where $\psi_\pm$ are the polarization degrees of freedom~\cite{Carusotto2013} with spin exchange symmetry.
Another experimental realization is the magnon condensation observed in microwave-pumped YIG films \cite{Demokritov2006,Nowik-Boltyk2012}, where the two modes arise from two minima in the band structure linked by inversion symmetry.  Both these cases realize our model~\eqref{eq:EoM_complex} for $N=2$ without fine-tuning, see~\ac{SM}~\cite{Note3}. Observable hallmarks are a first-order transition into the Van der Pol phase, and a universally divergent effective temperature, experimentally accessible through measuring the non-thermal mode occupation~\cite{Note3},  with an exponent $\eta-\text{Re}(\eta')$~\cite{Young2020,Sieberer2014} at the second-order transition. 

A striking aspect of our findings is that equilibrium fixed points can be unstable against any nonequilibrium perturbation. The only parallel example in the absence of conservation law that we are aware of is the Kardar-Parisi-Zhang equation in dimensions $d\leq 2$~\cite{Kardar1986}, characterizing a gapless nonequilibrium phase instead of a critical point.
Identifying the general principles behind destabilizing equilibrium fixed points in favor of nonequilibrium ones is an intriguing direction for future research -- such a mechanism would enable strong universal nonequilibrium effects under near equilibrium conditions. In turn, this can pave the way to genuine nonequilibrium scenarios in solid state, where usually nonequilibrium perturbations are leveled out by the opposite phenomenon, rapid thermalization. 

\begin{acknowledgments}

\paragraph{Acknowledgments--}
 We thank J. Lang, M. Tsitsishvili and B. Delamotte for useful discussions. We acknowledge support by the Deutsche Forschungsgemeinschaft (DFG, German Research Foundation) CRC 1238 project C04 number 277146847. 
 \end{acknowledgments}
 \FloatBarrier
%


\clearpage

\onecolumngrid
\setcounter{equation}{0}
\setcounter{figure}{0}
\setcounter{table}{0}
\setcounter{page}{1}
\makeatletter
\renewcommand{\theequation}{S\arabic{equation}}
\renewcommand{\thefigure}{S\arabic{figure}}
\renewcommand{\thetable}{S\arabic{table}}
\renewcommand{\bibnumfmt}[1]{[S#1]}
\renewcommand{\citenumfont}[1]{S#1}
\acresetall

\begin{center}
\textbf{\Large Supplemental Material: Nonequilibrium criticality at the onset of time-crystalline order}

\end{center}

In this Supplemental material, we present details about the field theoretical techniques used to study the stochastic Langevin equations. We concentrate on the complex Langevin equation
\begin{align}
\label{eq:EoM_complexApp}
    (\partial_t -Z\nabla^2  + \gamma)\boldsymbol{\psi} + \frac{g}{2} (\boldsymbol{\psi} \cdot \boldsymbol{\psi}^{*}) \boldsymbol{\psi} + \kappa  (\boldsymbol{\psi} \cdot \boldsymbol{\psi})\boldsymbol{\psi}^*+ \boldsymbol{\xi} =0,
\end{align}
with $Z= Z_d+i Z_c$, $g=i g_d+i g_c$ and $\kappa'=\kappa_d+i\kappa_c$, i.e., Eq.~\eqref{eq:EoM_complex} up to the change $i \to -i$ that we perform for convenience. $\boldsymbol{\xi} \in \mathbb{C}^N$ is a Gaussian noise with zero mean, variance $\langle \xi_i(\boldsymbol{x},t)\xi^*_j(\boldsymbol{x},t)\rangle=D\delta_{i,j}\delta(\boldsymbol{x}-\boldsymbol{x'})\delta(t-t')$, and $\langle \xi_i(\boldsymbol{x},t)\xi_j(\boldsymbol{x},t)\rangle=0$. The Langevin equation~\eqref{eq:EoM_12} is fully equivalent to it and gives the dynamics of the real and imaginary parts of the fields, $\boldsymbol{\psi}=\boldsymbol{\chi}_1+i \boldsymbol{\chi}_2$. 
This equation can be equivalently recast as
\begin{align}
\label{eq:Ham_form}
    \partial_t\boldsymbol{\psi}+ \frac{\delta H_d}{\delta\boldsymbol{\psi}^*}+i \frac{\delta H_c}{\delta \boldsymbol{\psi}^*}+ \boldsymbol{\xi}=0, \quad
   \text{with } H_a = \boldsymbol{\psi}^*\cdot(-Z_a\nabla^2 + \gamma_a)\boldsymbol{\psi} + \frac{g_a}{4} (\boldsymbol{\psi} \cdot \boldsymbol{\psi}^{*})^2 + \frac{\kappa_a}{2}(\boldsymbol{\psi} \cdot \boldsymbol{\psi})(\boldsymbol{\psi}^* \cdot \boldsymbol{\psi}^*), \quad a \in \{c,d\},
\end{align}
and $\gamma_c=0$, $\gamma_d=\gamma$. To make the link with the main text, one can also use $\rho=\boldsymbol{\psi} \cdot \boldsymbol{\psi}^{*}$ and $\tau=(\boldsymbol{\psi} \cdot \boldsymbol{\psi})(\boldsymbol{\psi}^* \cdot \boldsymbol{\psi}^*)/4$. 

In the absence of noise, this equation of motion is solved for configurations that verify $\delta H_d/\delta\boldsymbol{\psi}=0$ i.e., by configurations that minimize $H_d$,  while the nonzero remaining part coming from $H_{c}$ can always be canceled by an oscillation of the form $\exp(i\Delta \omega t)$ (for spatially homogeneous field configurations)~\cite{Taeuber2014a,Sieberer2016}. This oscillation corresponds to a redefinition of the original finite frequency scale, $\omega_0 \to \omega_0+ \Delta\omega$, see Eq.~\eqref{eq:defphi12}. Indeed, $\boldsymbol\psi(\boldsymbol{x},t) \to \exp(i\Delta\omega t)\boldsymbol\psi(\boldsymbol{x},t)$ translates to $(\boldsymbol{\chi}_1,\boldsymbol{\chi}_2)^T \to R(\Delta\omega t)(\boldsymbol{\chi}_1,\boldsymbol{\chi}_2)^T=R((\omega_0+\Delta\omega)t) (\boldsymbol{\phi},\boldsymbol\Pi)^T$. The stable states can thus be found as in an equilibrium problem, despite the nonequilibrium nature. 

Explicitly, for $\gamma>0$, the stable state is $\boldsymbol\psi_S=\langle\boldsymbol{\psi}\rangle=0$, while for $\gamma<0$, it has $\rho \neq 0$. For $\kappa_d>0$, the stable state has $\rho\neq 0, \tau= 0$, i.e., $\boldsymbol{\psi}\cdot \boldsymbol{\psi}=0$ or equivalently $\boldsymbol{\chi}_1 \perp \boldsymbol{\chi}_2$. This corresponds to the rotating phase. The equation of motion is, for example, solved by $\boldsymbol{\psi}_S(t)=\exp(i\Delta\omega t)\sqrt{-\gamma/g_d}(1,i,0,\dots,0)^T$ with $\Delta\omega=-\gamma g_c/g_d$. 
For $\kappa_d<0$, the stable configurations have $\tau\neq 0 $, e.g., $\boldsymbol{\psi}_S(t)=\exp(i\Delta\omega t)\sqrt{-2\gamma/(g_d+\kappa_d)}(1,0,\dots,0)^T$ with $\Delta\omega=-\gamma (g_c+\kappa_c)/(g_d+\kappa_d)$. This corresponds to the Van der Pol phase. 

\section{1. Functional integral formalism}
\subsection{a. MSRJD construction}

Using the \ac{MSRJD} construction~\cite{Martin1973,Janssen1976,Dominicis1976}, Eq.~\eqref{eq:EoM_complexApp} corresponds to the following functional integral~\cite{Taeuber2014}:
\begin{align}
\label{eq:zfunctionalint}
    Z[\boldsymbol{j},\tilde{\boldsymbol{j}}]=\int\mathcal{D}\boldsymbol{\psi}\mathcal{D}\tilde{\boldsymbol{\psi}} e^{-S[\boldsymbol{\psi},\tilde{\boldsymbol{\psi}}]+\int_{\boldsymbol{x},t}\tilde{\boldsymbol{j}}^\dag\boldsymbol{\psi}+\boldsymbol{j}^\dag\tilde{\boldsymbol{\psi}}},
\end{align}
where the action is given by, using $\int_{\boldsymbol{x},t} = \int d^d\boldsymbol{x} \, d t$,
\begin{align}
\label{eq:Action}
    S=& \int_{\boldsymbol{x},t} \tilde{ \boldsymbol{\psi}}^* \cdot (\partial_t \boldsymbol{\psi} -(1+ i r_{K,B})\nabla^2 \boldsymbol{\psi} + \gamma_{B}\boldsymbol{\psi}) +\mathrm{c.c.} - 2 \tilde{\boldsymbol{\psi}}^* \cdot \tilde{\boldsymbol{\psi}} + \frac{g_B}{2}(\tilde {\boldsymbol{\psi}}^* \cdot \boldsymbol{\psi}) (\boldsymbol{\psi}^* \cdot \boldsymbol{\psi})  +\frac{\kappa_B}{2}(\tilde {\boldsymbol{\psi}}^* \cdot \boldsymbol{\psi}^*)( \boldsymbol{\psi} \cdot \boldsymbol{\psi})  +\mathrm{c.c.}
\end{align}
In comparison with Eq.~\eqref{eq:EoM_complexApp}, we set $Z_d=D=1$ , and thus $r_{K,B}=Z_c/Z_d=Z_c$, since these coefficients can be absorbed into a redefinition of the fields. We include a $B$ index to all parameters of the action here, to indicate that they are bare microscopic quantities. The \ac{MSRJD} procedure can be applied to the original Langevin equation~\eqref{eq:EoMbegin} of the main text that involve real fields similarly, see~\cite{zelle23}. 

The functional integral $\ln Z[\boldsymbol{j},\tilde{\boldsymbol{j}}]$ can be used to generate all noise-averaged connected correlation and response functions of the problem. In particular, absent symmetry breaking, the two-point (retarded) response function and correlation functions are
\begin{align}
    \chi^R_{ij}(q,\omega)=\frac{\delta^2 \ln Z}{\delta \tilde {j}_i(q,\omega)\delta j_j^*(q,\omega)}\Big|_{\boldsymbol{j}=\tilde{\boldsymbol{j}} = 0}\equiv G^R(q,\omega) \delta_{ij}, \quad 
    \mathcal{C}_{ij}(q,\omega)=\frac{\delta^2 \ln Z}{\delta \tilde j_i(q,\omega)\delta \tilde {j}_j^*(q,\omega)}\Big|_{\boldsymbol{j}=\tilde{\boldsymbol{j}} = 0}\equiv G^K(q,\omega)\delta_{ij}.
\end{align}
We used time and space translation invariance to obtain diagonal propagators in Fourier space. The time-crystal phases break time translation invariance, but it is recovered in the effective $O(N) \times SO(2)$ theory, where the finite oscillating scale is implicit. The two-point functions can be collected into a matrix in Fourier space $\mathcal G(\boldsymbol{q},\omega)$, which absent symmetry breaking, is a $2\times2$ matrix in Keldysh space $(\boldsymbol{\psi},\tilde{\boldsymbol{\psi}})$,
\begin{align}
\label{eq:propagators}
    \mathcal{G}(\boldsymbol{q},\omega)=\begin{pmatrix}
        G^K(\boldsymbol{q},\omega) & G^R(\boldsymbol{q},\omega)\\
        G^A(\boldsymbol{q},\omega) & 0
    \end{pmatrix}, \quad \text{}
\end{align}
where $G^A = (G^R)^*$, $G^K = 2 D | G^R|^2$. At the Gaussian level, we have, from the action~\eqref{eq:Action}, 
\begin{align}
\label{eq:GrMF}
G^R(\boldsymbol{q},\omega)= \frac{1}{-i\omega + q^2(1+ i r_{K,B})+ \gamma_B}.
\end{align}

To ease the analysis, it is convenient to work with the effective action $\Gamma[\boldsymbol{\Psi},\tilde{ \boldsymbol{\Psi}}]$, with $\boldsymbol{\Psi}=\langle \boldsymbol{\psi} \rangle$ and $\tilde{\boldsymbol{\Psi}}=\langle \tilde{\boldsymbol{\psi}} \rangle$, defined as the Legendre transform of $\ln Z[\boldsymbol{j},\tilde{\boldsymbol{j}}]$~\cite{ZinnJustin,Taeuber2014},
\begin{equation}
    \Gamma[\boldsymbol{\Psi},\tilde{\boldsymbol{\Psi}}] = - \ln Z[\boldsymbol{j},\tilde{\boldsymbol{j}}]  +\int_{x,t}\tilde{\boldsymbol{j}}^\dag\boldsymbol{\Psi}+\boldsymbol{j}^\dag\tilde{\boldsymbol{\Psi}},\quad \textrm{where }  \quad \boldsymbol{j}= \frac{\delta \Gamma}{ \delta \tilde{\boldsymbol{\Psi}}^* }, \ \tilde{\boldsymbol{j}}= \frac{\delta \Gamma}{ \delta \boldsymbol{\Psi}^* }.
\end{equation}
The effective action is beneficial conceptually since it can be seen as a renormalized version of the action upon including the effect of interactions. It is also technically convenient because it is the generating functional of one-particle irreducible vertices, for which the perturbative computations done below are efficiently performed~\cite{ZinnJustin,Taeuber2014}.

\subsection{b. Thermal symmetry}

The MSRJD representation is also well suited to detect whether the system is in thermal equilibrium or not. In thermal equilibrium, all $n$-point correlation and response functions obey \ac{fdr}. In the field theoretical formalism, the presence of the \ac{fdr} are equivalent to the existence of a symmetry of the \ac{MSRJD} action and effective action~\cite{Janssen1976,Bausch1976,Andreanov2006,Aron2010,Sieberer2016,Gyoergyi1992,Aron2016}.

For a complex field, the thermal symmetry is given by~\cite{Sieberer2016}
\begin{equation}
\label{eq:ThermalSymm}
\begin{split}
    &\boldsymbol\Psi(\boldsymbol{x},t)\rightarrow \boldsymbol\Psi(\boldsymbol{x},-t)^*, \quad \tilde{\boldsymbol\Psi}(\boldsymbol{x},t)\rightarrow \tilde{\boldsymbol\Psi}(\boldsymbol{x},-t)^*+\frac{1}{2 T}\partial_t\boldsymbol\Psi(\boldsymbol{x},-t)^*,
\end{split}    
\end{equation}
where $T$ denotes the equilibrium temperature. Evidently, the thermal symmetry of the original fields is broken in the limit-cycle phases~\cite{zelle23}. This propagates to the effective theory, and Eq.~\eqref{eq:ThermalSymm} is not a symmetry of the action~\eqref{eq:Action} either. However, one can allow for a more general thermal symmetry in the presence of coherent and dissipative dynamics~\cite{Sieberer2014},
\begin{equation}
\label{eq:ThermalSymm2}
\begin{split}
    &\boldsymbol\Psi(\boldsymbol{x},t)\rightarrow \boldsymbol\Psi(\boldsymbol{x},-t)^*, \quad \tilde{\boldsymbol\Psi}_m(\boldsymbol{x},t)\rightarrow \tilde{\boldsymbol\Psi}_m(\boldsymbol{x},-t)^*+\frac{1}{2 T}\partial_t\boldsymbol\Psi(\boldsymbol{x},-t)^*, \text{ where} \quad \tilde{\boldsymbol\Psi}_m(\boldsymbol{x},t)=(1+i b)\tilde{\boldsymbol\Psi}(\boldsymbol{x},t),
\end{split}    
\end{equation}
with $b$ an additional parameter.
The action~Eq.~\eqref{eq:Action} is now symmetric under Eq.~\eqref{eq:ThermalSymm2} if and only if $ H_c= b H_d$ (which fixes the parameter $b$) with temperature $T=D/2=1/2$ in our units. This implies a fixed ratio between the real and imaginary parts of the couplings, $r_K=g_c/g_d=\kappa_c/\kappa_d$.  There is no condition between the real and imaginary parts of $\gamma=i\gamma_d+\gamma_c$ because we can always shift the value of $\gamma_c \to \gamma_c+\Delta\omega$ via $\boldsymbol\psi(t) \to \exp(i\Delta\omega t)\boldsymbol\psi(t)$, i.e., a redefinition of the finite frequency parameter $\omega_0$ as discussed above. (This means that the corresponding effective thermal behavior is found in a rotating frame.)

The presence of this symmetry can be rationalized by noting that, if $ H_c= b H_d$, we can rewrite the Langevin equation~\eqref{eq:EoM_complex} as
\begin{align}
   \frac{ \partial_t \boldsymbol{\psi}}{1-i b } + \frac{\delta H_d}{\delta\boldsymbol{\psi}^*}+ \frac{\boldsymbol{\xi}}{1-i b} =0, 
\end{align}
and we in fact recover a purely conservative Hamiltonian dynamics that describe thermal equilibrium with a noise (i.e., $\boldsymbol{\tilde \psi}$) rescaled by $(1-i b)$, in agreement with Eq.~\eqref{eq:ThermalSymm2}. 


Limit cycles are nonequilibrium phases. But this does not preclude the possibility of an \textit{effective} equilibrium (in a rotating frame) at the transition, even if the starting microscopic model breaks equilibrium conditions. In particular, this happens in the $N=1$ case of the model we are considering here.  

\section{I. Goldstone theorem and dispersions}
\subsection{a. Spontaneous breaking of time-translation symmetry and Goldstone theorem}

In this section, we analyze the original equation~\eqref{eq:EoMbegin}, and discuss the Goldstone modes associated with time-translation breaking.
A continuous symmetry with generators $T_{ij}$, in its infinitesimal version, acts on the field, $\boldsymbol{\phi}$, as $\phi_i \to (1+ \epsilon T_{ij})\phi_j$. The Goldstone theorem reads as, for a space independent order parameter $\boldsymbol{\phi}_S(t)=\langle \boldsymbol{\phi}(\boldsymbol{x},t)\rangle$,
\begin{align}
    \sum_{i,j}\int_{t'} (\left.G^R\right.^{-1})_{ki}(\boldsymbol{q}=0,t,t')T_{ij} \phi_{S,j}(t') =0.
\end{align}
For a broken generator, $\sum_{j}T_{ij}\boldsymbol{\phi}_{S,j}(t')$ is nonzero, and $G^R$ necessarily has a (eigen-)mode with a vanishing dispersion at $\boldsymbol{q}=0$, i.e., a Goldstone mode.

It is shown in~\cite{zelle23} that the Goldstone theorem leads to $2N-3$ Goldstone modes associated with the breaking of the $O(N)$ symmetry in the rotating phase. The breaking of the $O(N)$ symmetry in the Van der Pol phase gives $N-1$ Goldstone modes, as in the usual static phase of $O(N)$ models.

The \ac{MSRJD} functional integral equivalent to the initial Langevin equation Eq.~\eqref{eq:EoMbegin} of the main text is invariant under infinitesimal time translations $t\to t+\epsilon,\, \boldsymbol{\phi}' \to \boldsymbol{\phi}+\epsilon\partial_t \boldsymbol{\phi}$ whose only generator is $T_{ij}=\delta_{ij}\partial_t$. We therefore get one Goldstone mode from its breaking whenever the order parameter is time-dependent,
\begin{align}
\label{eq:Gold_T_tt}
    \sum_{i}\int_{t'} (\left.G^R\right.^{-1})_{ki}(\boldsymbol{q}=0,t,t')\partial_t'\phi_{S,i}(t') =0.
\end{align}

In the Van der Pol phase, this gives one Goldstone mode that arises solely from this spontaneous breaking of time-translation symmetry. 
On the contrary, in the rotating phase the associated Goldstone mode is equivalent to the one arising from the rotation along the limit cycle~\cite{zelle23} since the Goldstone theorem applied for the associated broken generator of $O(N)$ leads to the same expression as the one obtained from~\eqref{eq:Gold_T_tt}.

All the Goldstone modes can be also obtained from the $O(N)\times SO(2)$ theory in a close manner to the equilibrium case. In particular, the Van der Pol phase corresponds to the breaking pattern $O(N)\times SO(2) \to O(N-1)$. There are again $N-1$ Goldstones arising from the breaking of $O(N)$, and one Goldstone from the breaking of the $SO(2)$ symmetry. In the rotating phases, the breaking pattern is $O(N)\times SO(2) \to O(N-2)\times SO(2)_d$ (the additional index in $SO(2)_d$ underlines that it differs from the original one~\cite{Delamotte2004}). There are  $2N-3$ Goldstones coming from the breaking of $O(N)$ to $O(N-2)$, while again the Goldstone of the broken $SO(2)$ is not an independent one.

\subsection{b. Dispersions of the Goldstone modes}

The effective $O(N)\times SO(2)$ theory allows us to recover effective time-independent dynamics for the Goldstone modes because we can now expand around time-independent solutions. In turn, we can get explicit dispersion relations $\omega(q)$ for the Goldstone modes. This is rationalized by the Floquet theorem, which tell us that there exist periodic functions $P(t)$ such that the linearized solutions of the equation of motion are of the form
\begin{align}
\label{eq:Floquet}
    \delta\boldsymbol{\phi}(\boldsymbol{q},t)=\boldsymbol{P}(t)\exp(-i \omega(q) t).
\end{align}
This confirms that we were able to work in the ``rotating frames'' around the two limit-cycle phases. We now derive the dispersion relations by specifying explicit forms for the real and imaginary parts of $\boldsymbol\psi=\boldsymbol{\chi}_1+i\boldsymbol{\chi}_2$ in the broken phase.

\paragraph{Rotating phase --} One of the possible choices is to parameterize the fields by introducing amplitude $\delta \rho_i(\boldsymbol{x},t)$ and angular fluctuations $\theta_{i,j}(\boldsymbol{x},t)$ (with $i \in \{1,2\}$ and $j \in \{3 \dots N\}$):
\begin{align}
    &\boldsymbol{\chi}_1(\boldsymbol{x},t)= \sqrt{\rho_0+\delta\rho_1}\exp[\theta_{1,1} T_{1,2}+\sum_{i=3}^{N} \theta_{1,i} T_{1,N}]\hat {\boldsymbol{e}}_1, \quad 
    \boldsymbol{\chi}_2(\boldsymbol{x},t)= \sqrt{\rho_0+\delta\rho_2}\exp[\theta_{2,1} T_{2,1}+\sum_{i=3}^{N} \theta_{2,i}T_{1,N}]\hat {\boldsymbol{e}}_2.
\end{align}

The amplitude modes are gapped in the broken phases and can be safely integrated out. In addition, because $\boldsymbol{\chi}_1$ and $\boldsymbol{\chi}_2$ are orthogonal, only rotations that keep their relative angle fixed lead to a soft mode. Therefore, the relative angle mode $\theta_{1,2}+\theta_{2,1}$ is also gapped, while $\theta_-=1/\sqrt{2}(\theta_{1,2}-\theta_{2,1})$ is gapless. Its linearized dynamics and dispersion relation are given by
\begin{align}
\label{eq:Goldstone_rot_new1app}
    &\partial_t \theta_- -Z_d\nabla^2 \theta_-  +\xi = 0 \implies \omega(q)= -i Z_d q^2. 
\end{align}
All the other $\theta$ modes are also gapless and coupled by pairs at the linear level:
\begin{align}
\label{eq:Goldstone_rot_new2app}
    &\partial_t \theta_{1,i}-Z_d\nabla^2 \theta_{1,i}+Z_c\nabla^2 \theta_{2,i}  +\xi = 0, \quad \partial_t \theta_{2,i}-Z_d\nabla^2 \theta_{2,i}-Z_c\nabla^2 \theta_{2,i}  +\xi = 0, \quad \forall i \in [3\dots N].
\end{align}
 Instead of purely overdamped motion, this leads to two modes, \(\omega_{\pm}(q)= -i Z_d q^2 \pm Z_c q^2\).

These modes were found in~\cite{zelle23} with an additional finite gap equal to the frequency of the order parameter $\omega_0$, whose absence here is due to the rotation~\eqref{eq:defphi12}. This can again be  rationalized in the Floquet language. Indeed, from Eq.~\eqref{eq:Floquet}, we can write
\begin{align}
    \delta\boldsymbol{\phi}(\boldsymbol{q},t)&=\boldsymbol{P}(t)\exp(-i\omega(q) t)=\boldsymbol{P}_2(t)\exp(-i\omega(q)t - i \omega_0 t),
\end{align}
where $\boldsymbol{P}_2(t)$ is still periodic with frequency $\omega_0$.
 Therefore, if a mode oscillates at $\pm \omega_0$ as the Goldstone modes in~\cite{zelle23}, it is possible to transfer the oscillation into $\boldsymbol{P}(t)$, and get a non-oscillating mode instead.

\paragraph{Van der Pol phase --}

The effective dynamics for the Goldstone modes are obtained by writing the fields as
\begin{align}
    &\boldsymbol{\chi}_1(\boldsymbol{x},t)= (\boldsymbol{\chi}_0+\sigma_1(\boldsymbol{x},t),\theta_2(\boldsymbol{x},t),\dots,\theta_{N}(\boldsymbol{x},t)), \quad    \boldsymbol{\chi}_2(\boldsymbol{x},t)= (\theta_1(\boldsymbol{x},t),\sigma_2(\boldsymbol{x},t),\dots \sigma_N(\boldsymbol{x},t) ).
\end{align}
All the $\sigma$ modes have a finite mass at the Gaussian level, while the $\theta$ modes are gapless with
\begin{align}
\label{eq:EoMGold_VdPapp}
    \partial_t \theta_i -Z_d\nabla^2 \theta_i+ \xi =0.
\end{align}

From the point of view of the original equation of motion~\eqref{eq:EoMbegin}, the effective $O(N)\times SO(2)$ theory describes the Van der Pol phase only to the lowest order in $\gamma$. In principle, the Van der Pol limit cycle can be obtained to any order in perturbation theory~\cite{Verhulst1996}, around which the Floquet exponents, or equivalently stated, the dispersion of the Goldstone modes can be extracted. We can expect the linearized Langevin equations to still be of the form \eqref{eq:EoMGold_VdPapp} based on symmetry breaking pattern considerations.

The Gaussian fluctuations of these Goldstone modes, $\langle \theta(\boldsymbol{x},t)^2\rangle \sim \int_{\boldsymbol{q},\omega}G^K_\theta(\boldsymbol{q},\omega) \sim \int d^d \boldsymbol{q}/q^2 $, diverge in $d\leq 2$. They thus destroy any order in two dimensions, and the Mermin-Wagner theorem applies. This is true even for $N=1$, where this is due solely to the Goldstone mode associated with the continuous time-translation symmetry. 

\section{II. Details about the RG procedure}
\subsection{1. Scaling hypothesis and dimensionless action}

To take into account interaction effects in perturbation theory, it is sufficient to parameterize the effective action as  $\Gamma= \Gamma_0+ \Gamma_{\mathrm{int}}$ with 
\begin{align}
    \Gamma_0 &= \int_{\boldsymbol{x},t} \tilde{\boldsymbol{\psi}}^* \cdot (Z_t^{-1} \partial_t -Z_x^{-1}\nabla^2  + \gamma)\boldsymbol{\psi} +\mathrm{c.c.} - 2 Z_D^{-1} \tilde {\boldsymbol{\psi}}^* \cdot \tilde{\boldsymbol{\psi}} ,\\
    \Gamma_{\mathrm{int}}&= \int_{\boldsymbol{x},t} \frac{g}{2}(\tilde {\boldsymbol{\psi}}^* \cdot \boldsymbol{\psi}) (\boldsymbol{\psi}^* \cdot \boldsymbol{\psi})  +\frac{\kappa}{2}(\tilde {\boldsymbol{\psi}}^* \cdot \boldsymbol{\psi}^*)( \boldsymbol{\psi} \cdot \boldsymbol{\psi})  +\mathrm{c.c.},
\end{align}
 where all parameters, except the real valued $Z_D$, are complex numbers. This form involves only relevant and marginal operators in four dimensions, following power counting arguments. We will determine the renormalization group equations, which describe how the effective couplings entering the effective action change upon variation of the momentum scale $\mu$ at which we define them.

The prefactor of the time derivative does not remain real upon coarse-graining, even if initialized as such on the microscopic scale. In order to restore the form of the bare action, it is therefore convenient to first rescale the response field to eliminate the $Z_t$ factor by introducing $\tilde {\boldsymbol{\psi}}'= (Z_t^{-1})^* \tilde{\boldsymbol{\psi}}$, 
\begin{align}
    \Gamma_0 &= \int_{\boldsymbol{x},t} \tilde{\boldsymbol{\psi}}'^* \cdot (\partial_t -\frac{Z_x^{-1}}{Z_t^{-1}}\nabla^2 + \frac{r}{Z_t^{-1}})\boldsymbol{\psi} +\mathrm{c.c.} - 2 \frac{Z_D^{-1}}{|Z_t|^{-2}} \tilde{\boldsymbol{\psi}}'^* \cdot \tilde{\boldsymbol{\psi}}'.
\end{align}
We define $Z_x^{-1}/Z_t^{-1} \equiv K \equiv K_d+ i K_c$. The $\gamma$ parameter does not either remain real, but its imaginary part can always be eliminated by a redefinition of the finite frequency, $\omega_0$, of the order parameter, see above.

To make the scale invariance of the fixed points explicit, the action has to be put in a dimensionless form by rescaling the fields, space, and time to absorb bare and anomalous dimensions,
\begin{align}
\label{eq:DimLess}
    \boldsymbol{x}_R = \mu \boldsymbol{x},\quad t_R = \mu^2 X^{-1} t ,  \quad \tilde {\boldsymbol{\psi}}_R = \mu^{-\frac{d+2}{2}} \sqrt{Z_{\tilde{\psi}}^{-1}}\tilde{\boldsymbol{\psi}}', \quad \boldsymbol{\psi}_R = \mu^{-\frac{d-2}{2}} \sqrt{Z_{\psi}^{-1}}\boldsymbol{\psi}.
\end{align}
The effective action is then dimensionless upon choosing 
\begin{equation}
   X=\frac{1}{K_d}, \quad  Z_{\tilde\psi}=\frac{Z_D}{|Z_t|^2 X}, \quad Z_{\psi}=Z_{\tilde\psi}^{-1},
\end{equation}
and reads as
\begin{align}
\label{eq:EffActDimless}
    \Gamma &= \int_{x_R,t_R} \tilde{\boldsymbol{\psi}}_R^* \cdot (\partial_t \boldsymbol{\psi}_R -(1+ i r_K)\nabla^2 \boldsymbol{\psi}_R +\gamma \tilde {\boldsymbol{\psi}}_R) +\mathrm{c.c.} - 2 \tilde {\boldsymbol{\psi}}_R^* \cdot \tilde{\boldsymbol{\psi}}_R+ \frac{\tilde g}{2}(\tilde {\boldsymbol{\psi}}_R^* \cdot \boldsymbol{\psi}_R) (\boldsymbol{\psi}_R^* \cdot \boldsymbol{\psi}_R)  +\frac{\tilde \kappa}{2}(\tilde {\boldsymbol{\psi}}_R^* \cdot \boldsymbol{\psi}_R^*)( \boldsymbol{\psi}_R \cdot \boldsymbol{\psi}_R)  +\mathrm{c.c.},
\end{align}
where
\begin{align}
\label{eq:DimLessCouplings}
    \tilde \gamma = \gamma X Z_t/\mu^2, \quad \tilde{g}= \mu^{d-4} g Z_\psi Z_t X,  \quad \tilde{\kappa}= \mu^{d-4} \kappa Z_\psi Z_t X.
\end{align}
The additional parameter $r_K=K_c/K_d$, which cannot be eliminated by  the rescaling,  describes the relative strength of coherent couplings vs. dissipative couplings in the two-point functions.

All critical exponents can then be defined in the standard way from the dynamical scaling hypothesis~\cite{Taeuber2014,Taeuber2014a},
\begin{align}
    G^K(\boldsymbol{q},t) &= q^{-2+ \eta} \tilde G^K(t q^z(1+ i q^{\eta-\eta_c} ), \gamma q^{-1/\nu})= q^{-2+ \eta} \tilde G^K(t q^z, i q^{\eta-\eta_c}, \gamma q^{-1/\nu}),\\  G^R(\boldsymbol{q},t) &= q^{-2+\eta'+z} \tilde G^R(t q^z, i q^{\eta-\eta_c}, \gamma q^{-1/\nu}), 
\end{align}
where the additional independent exponent $\eta_c$ allows for a possibly different scaling between coherent and dissipative couplings~\cite{Taeuber2014a}. It is associated with the scaling behavior of the parameter $r_K \sim \mu^{\eta-\eta_c}$. At equilibrium, the \ac{fdr} imply $\eta=\eta'$, and there is one less independent critical exponent. This ansatz encompasses the Gaussian case Eq.~\eqref{eq:GrMF}, for which $\nu=1/2$, $z=2$ and $\eta=\eta'=\eta_c=0$.

Beyond mean field, we can use the dimensionless effective action~\eqref{eq:EffActDimless} to write 
\begin{align}
\label{eq:dimless1}
    G^K(\boldsymbol{q},t)&=\langle \psi_i^* (\boldsymbol{q},t) \psi_i(\boldsymbol{q},0)\rangle = \mu^{-2} Z_\psi \langle \psi_{R,i}^* (\boldsymbol{q},t) \psi_{R,i}(\boldsymbol{q},0)\rangle = \mu^{-2} Z_\psi \tilde G^K(\boldsymbol{q}_R=q/\mu, t_R= t\mu^2 X^{-1}, i r_K,\tilde \gamma ), \\
    \label{eq:dimless2}
     G^R(\boldsymbol{q},t)&=\langle \tilde {\psi}^*_i (\boldsymbol{q},t) \psi_i(\boldsymbol{q},0)\rangle =  Z_t \langle \tilde {\psi}^*_{R,i}(\boldsymbol{q},t) \psi_{R,i}(\boldsymbol{q},0)\rangle = Z_t \tilde G^R(\boldsymbol{q}_R, t_R, i r_K, \tilde \gamma ).
\end{align}
In these expressions, the RG parameter $\mu$ plays a similar role as $q$ since it is also a momentum scale. Based on the scaling hypothesis, we can then use the standard RG matching procedure, $q \sim \mu$, to read off scaling from the RG flow. We therefore have 
\begin{align}
    \nu^{-1}= -\mu \partial_\mu \ln(\tilde\gamma), \quad \eta=\mu\partial_\mu \ln Z_\psi, \quad z=2-\mu\partial_\mu \ln X, \quad \eta'= \mu \partial_\mu \ln(Z_t)+\mu\partial_\mu \ln X, \quad \eta_c = \eta-\mu\partial_\mu \ln(r_K).
\end{align}

In equilibrium, the \ac{fdr} for the two-point response and correlation functions reads as
\begin{align}
\label{eq:Tfdr}
    G^R(\boldsymbol{q},t)-G^A(\boldsymbol{q},t)= -\frac{1}{2 T} \partial_t G^K(\boldsymbol{q},t).
\end{align}
This can be taken as the definition of the temperature via $T=-\partial_t G^K(\boldsymbol{q},t)/ 2 G^R(\boldsymbol{q},t)$ for $t>0$. Out-of-equilibrium, this \ac{fdr} does not hold anymore. If one insists on the definition~\eqref{eq:Tfdr}, this formally shows up as a scale dependent temperature $T_\mu$ . Specifically, we find from~\eqref{eq:dimless1} and~\eqref{eq:dimless2} that, in the scaling regime where $q\lesssim \mu$ and $t \lesssim \mu^{-z}$,
\begin{align}
    T_\mu = \left|\frac{ \partial_t G^K(\boldsymbol{q},t)}{ 2 G^R(\boldsymbol{q},t)}\right| \sim |  Z_\psi Z_t^{-1} \mu^{-2+z}| \sim \mu^{\textrm{Re}(\eta-\eta')}. \label{eq:temp}
\end{align}
At equilibrium $\eta=\eta'$ and a true (scale free) temperature is found. Reciprocally, we see that a nonequilibrium situation can possibly be detected by measuring the ratio between response and correlation functions, or equivalently the distribution of excitations, in a time or space resolved way.

\subsection{2. Perturbation theory}
\begin{figure}
    \centering
    \subfloat[][Retarded propagator]{\label{fig:Gr}
\(G^R(\boldsymbol{q},\omega)= \includegraphics[valign=c]{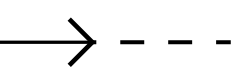}\)
}
\hfil
    \subfloat[][Advanced propagator]{\label{fig:Ga}
\(G^A(\boldsymbol{q},\omega)= \includegraphics[valign=c]{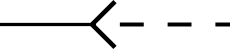}\)
}
\hfil
\subfloat[][Keldysh propagator]{\label{fig:Gk}
   \(G^K(\boldsymbol{q},\omega)= \includegraphics[valign=c]{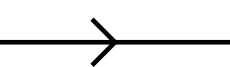}\)
    }
\vfil
\subfloat[][Four-point vertex]{\label{fig:vertex}
   \(\frac{g}{2}(\delta_{a,c}\delta_{b,d} +\delta_{a,d}\delta_{b,c})+\kappa \delta_{ab}\delta_{cd}= \includegraphics[valign=c]{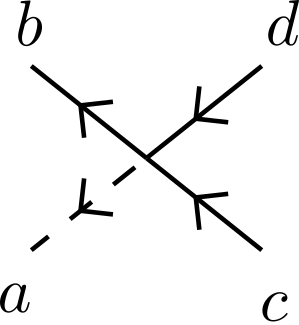}\)
    }
    \caption{Diagrammatic representation of the elements entering perturbation theory. The propagators are defined in Eq.~\eqref{eq:propagators}. The complex conjugated vertex, obtained by reverting the direction of the arrows in (d), is not shown.}
    \label{fig:Elt_PT}
\end{figure}
In the spirit of renormalized perturbation theory, it is convenient to work directly in renormalized dimensionless variables by first performing the transformation~\eqref{eq:DimLess}, and then to define the renormalized quantities by introducing multiplicative $Z$ factors via
\begin{equation}
    \gamma_B = Z_\gamma'  \tilde{\gamma}, \quad r_{K,B}= Z_{r_K} r_K, \quad g_{B}= Z_g' \tilde{g}, \quad \kappa_{B}= Z_\kappa' \tilde{\kappa}.
\end{equation}
In terms of these variables, the action reads
\begin{align}
    S = \int_{x,t} \tilde {\boldsymbol{\psi}}^* \cdot(Z_{t}\partial_t  - Z_{t}X(1+i Z_{K} r_{K})\nabla^2 + Z_{\gamma}\tilde\gamma)\boldsymbol{\psi} +\mathrm{c.c.} - 2 Z_{D} \tilde {\boldsymbol{\psi}}^* \cdot \tilde{\boldsymbol{\psi}},
\end{align}
where we additionally define $Z_\gamma= \mu^{-2} Z_t X Z_\gamma'$,  $Z_g= \mu^{d-4}Z_g' Z_t Z_\psi^2$ using Eqs.~\eqref{eq:DimLessCouplings}. We dropped all $R$ indices for simplicity.
All counterterms are of the form $Z_a= 1+\delta_a$, where the $\delta_a$, defined through these relations, have an expansion in terms of the coupling constants that starts at least at order one. 

Defining $Z_x=Z_{t}X(1+ i Z_{r_{K}}r_{K})= Z_{d}+ i Z_{c}r_{K}$, we can alternatively use the counterterms $Z_{d}=1+\delta_{d}$, $Z_{c}=1+\delta_{c}$. We will need the following relation
\begin{equation}
\label{eq:LinkCounter}
\delta_{r_{K}} = \frac{\delta_{c}}{r_{K}}-\delta_{d}-\text{Im}(\delta_{t})\frac{1+r_{K}^{2}}{r_{K}},  
\end{equation} 
valid at second-order in the interactions.
The action can then be written as $S= S_{0}+ S_{P}$ with 
\begin{align}
    S_{0}=& \int_{x,t} \tilde {\boldsymbol{\psi}}^* \cdot (\partial_t  -(1 + i r_K)\nabla^2  + \tilde \gamma )\boldsymbol{\psi} +\mathrm{c.c.} - 2 \tilde {\boldsymbol{\psi}}^* \cdot \tilde{\boldsymbol{\psi}} \\
     S_P =&  \int_{x,t} \tilde {\boldsymbol{\psi}}^* \cdot (\delta_t -\delta_x\nabla^2 + \delta_\gamma\tilde \gamma) \boldsymbol{\psi}- 2 \delta_D\tilde {\boldsymbol{\psi}}^* \cdot \tilde{\boldsymbol{\psi}}  \nonumber \\ 
     &+\frac{\tilde g(1+\delta_g)}{2}(\tilde {\boldsymbol{\psi}}^* \cdot \boldsymbol{\psi}) (\boldsymbol{\psi}^* \cdot \boldsymbol{\psi})  +\frac{\tilde \kappa(1+\delta_\kappa)}{2}(\tilde {\boldsymbol{\psi}}^* \cdot \boldsymbol{\psi}^*)( \boldsymbol{\psi} \cdot \boldsymbol{\psi})  +\mathrm{c.c.},    
\end{align}
and the loop corrections can be computed by treating $S_{P}$ as a perturbation. We use dimensional regularization with $\epsilon=4-d$ and the minimal subtraction scheme where only the poles in $1/\epsilon$ are incorporated into the $Z$ factors, see e.g.,~\cite{ZinnJustin,Kleinert2001,Taeuber2014}. The elements entering perturbation theory are displayed in Fig.~\ref{fig:Elt_PT}.

\begin{figure}
    \centering 
    \subfloat[\label{fig:tadpole}]{%
  \includegraphics{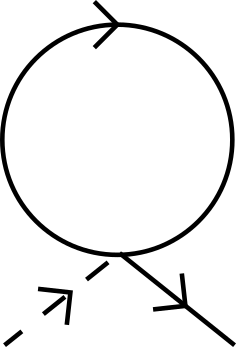}%
}\hfil
    \subfloat[\label{fig:sunset1}]{%
  \includegraphics{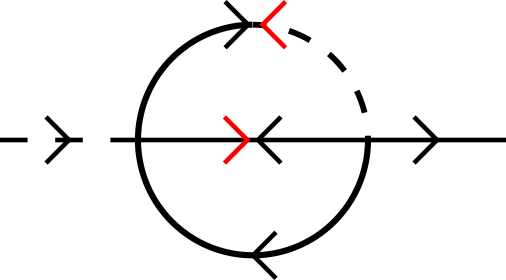}%
}
\hfil
    \subfloat[\label{fig:sunset2}]{%
  \includegraphics{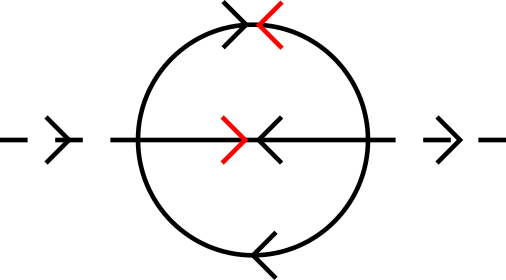}%
}\vfil
    \subfloat[\label{fig:vertexCorr}]{%
  \includegraphics{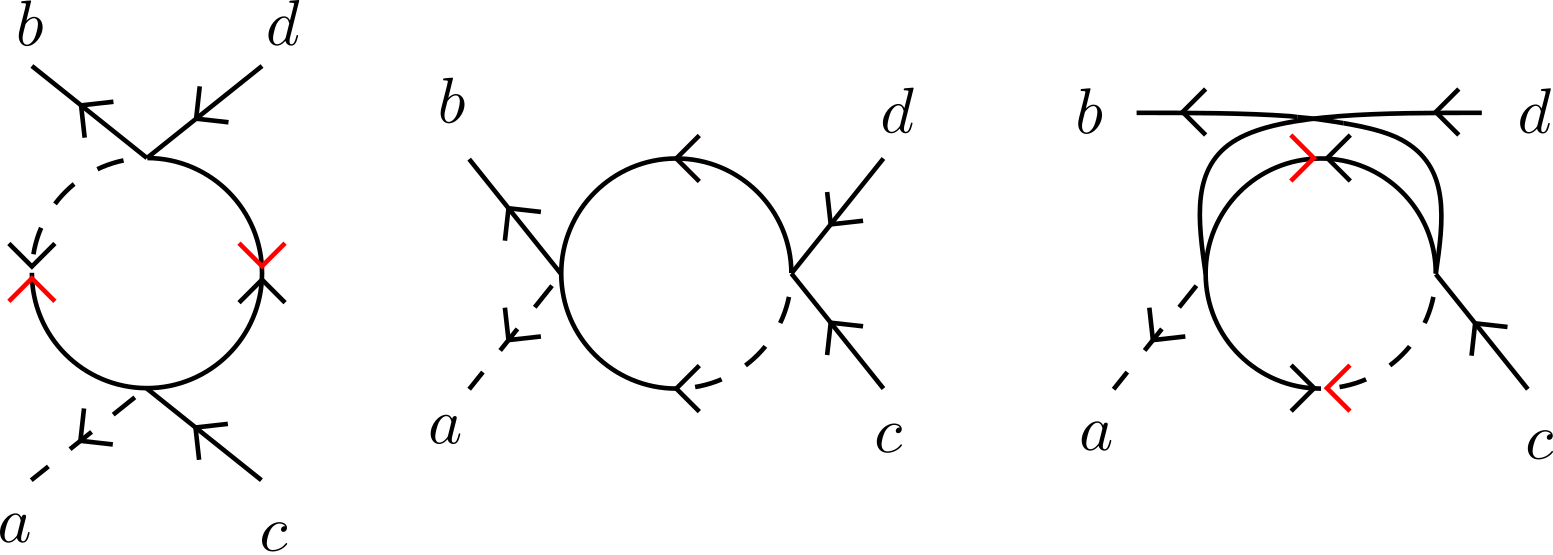}%
}
\caption{Loop diagrams considered in the text. The first three graphs renormalize the self-energies. (a) and (b) correct the retarded part of the action $\Gamma^{(2)}_{\tilde\psi_a^*\psi_a}$, while (c) corrects the noise part $\Gamma^{(2)}_{\tilde\psi_a^*\tilde \psi_a}$. The one-loop diagrams (d) renormalize the interaction. The red arrows indicate that diagrams with both arrow directions have to be considered.}
    \label{fig:2loop2pointFunc}
\end{figure}
\subsubsection{a. One-loop}
 We use the following notation to denote functional derivatives $\delta^2\Gamma / \delta\tilde\psi_a^*(\boldsymbol{p},\omega) \psi_b(\boldsymbol{p},\omega) = \Gamma^{(2)}_{\tilde\psi_a^* \psi_b } (p,\omega)$ and its generalizations to higher order. The only one-loop contribution to the self-energy is the tadpole diagram Fig.~\ref{fig:tadpole}, which gives
\begin{align}
\label{eq:1l}
    \Gamma^{(2)}_{\tilde\psi^*_a \psi_a}(p,\omega)= -i\omega+ p^2+ \tilde\gamma + \delta_{\gamma}\tilde\gamma + (\tilde g \frac{N+1}{2}+ \tilde \kappa)\int_{\boldsymbol{q},\omega} G^K(\boldsymbol{q},\omega) + O(I^2), 
\end{align}
where $O(I^2)\equiv  O(g^2,\kappa^2, g\kappa)$ and $\int_{{\boldsymbol{q},\omega}}=\int d^d\boldsymbol{q}d\omega/(2 \pi)^{d+1}$.
The frequency integrals can be performed in all loops without generating any divergence, while the divergent parts of the remaining momentum integrals are obtained in the standard way. All the poles of the integrals that we will need can also be found in~\cite{Risler2005,Taeuber2014a}. The divergences are the same as in the $N=1$ case, but the additional structure of the interactions for $N>1$ induces more complex flow equations as well as different prefactors of the loops. This rationalizes why we can get novel fixed points, and a novel universality class. The pole in Eq.~\eqref{eq:1l} is given by
\begin{align}
    \int_{\boldsymbol{q},\omega} G^K(\boldsymbol{q},\omega) = \int_{\boldsymbol{q}} \frac{1}{q^2 + \tilde\gamma} = \frac{2}{(4 \pi)^2 \epsilon} \tilde\gamma + O(\epsilon^0).
\end{align}

The renormalization of the four-point functions is given by the diagrams in Fig.~\ref{fig:vertexCorr} evaluated at zero momenta and frequencies. We obtain
\begin{align}
\begin{split}
    \Gamma^{(4)}_{\tilde\psi_a^* \psi_b^* \psi_c \psi_d}=& \left(\tilde g+\delta_g \tilde g - \frac{1}{(4\pi)^2\epsilon }\left[\frac{(N+3)\tilde g}{2}(\tilde g+\tilde g^*) +2 \tilde\kappa\tilde \kappa^* +2 \tilde g\tilde\kappa+ g^* \tilde\kappa+\tilde g \tilde\kappa^* + \frac{1}{1+i r_K} \tilde g^2\right]\right) \frac{1}{2}(\delta_{a,c}\delta_{b,d} +\delta_{a,d}\delta_{b,c}  )\\
   &+\left(\tilde\kappa+\delta_\kappa \tilde\kappa- \frac{1}{(4\pi)^2\epsilon }\left[2 \tilde g\tilde\kappa+ \tilde g^* \tilde\kappa+\tilde g \tilde\kappa^* +\frac{1}{1+i r_K}(N \tilde\kappa^2+2\tilde\kappa g) \right]\right) \delta_{ab}\delta_{cd}+ O(I^2, \epsilon^0).
\end{split}
\end{align}
By absorbing the poles in $\epsilon$ in the countertems, we have at one-loop order,
\begin{align}
    Z_\gamma &= 1 + \frac{2}{\epsilon} (\tilde g \frac{N+1}{2}+ \tilde\kappa)\tilde\gamma +O(I^2),\\
    Z_g &= 1 + \frac{1}{\epsilon }\left[\frac{(N+3)}{2}(\tilde g^2+\tilde g \tilde g^*) +2 \tilde\kappa\tilde \kappa^* +2 \tilde g\tilde\kappa+ \tilde g^* \tilde \kappa+\tilde g \tilde \kappa^* + \frac{1}{1+i r_K}\tilde g^2\right]+O(I^2), \\
    Z_\kappa &= 1+\frac{1}{\epsilon }\left[2 \tilde g\tilde \kappa+ \tilde g^* \tilde \kappa+\tilde g\tilde \kappa^* +\frac{1}{1+i r_K}(N \tilde\kappa^2+2\tilde\kappa\tilde g) \right]+O(I^2).
\end{align}
We redefine the couplings from now on via $g \to g (4\pi)^2$ and $\kappa \to \kappa (4\pi)^2$ to absorb all factors of $4\pi$. The RG equations are then obtained by using the fact that the bare quantities are independent of the RG scale $\mu$. For example, using $Z_g \tilde g \mu^\epsilon= g_B$ at one-loop (where $Z_g=Z_g'$), we get $\mu\partial_\mu \tilde g = -\epsilon- \mu\partial_\mu Z_g /Z_g $.  Taking real and imaginary part of the ensuing equations, we finally obtain 
\begin{align}
    \mu\partial_\mu \tilde \gamma &= -2 \tilde{\gamma} +\frac{\tilde{g}(N+1)+2 \tilde{\kappa}}{2}\tilde \gamma ,\\
    \mu\partial_\mu \tilde{g}_d &= -\epsilon \tilde{g}_d+ \left(
    \frac{r_K \tilde{g}_c \tilde{g}_d+\frac{1}{2}
   \left(\tilde{g}_d^2-\tilde{g}_c^2\right)}{r_K^2+1}+2 \tilde{g}_d
   \tilde{\kappa }_d+\frac{1}{2} (N+3) \tilde{g}_d^2+2 \tilde{\kappa
   }_d^2\right),\\
    \mu\partial_\mu \tilde{g}_c &= -\epsilon \tilde{g}_c+ \left(
    \tilde{\kappa }_c \tilde{g}_d+\tilde{g}_c \tilde{\kappa }_d+\frac{1}{2}(N+3) \tilde{g}_c \tilde{g}_d+\frac{\frac{1}{2} r_K\left(\tilde{g}_c^2-\tilde{g}_d^2\right)+\tilde{g}_c\tilde{g}_d}{r_K^2+1}+2 \tilde{\kappa }_c \tilde{\kappa }_d \right),\\
    \mu\partial_\mu \tilde{\kappa}_d &= -\epsilon \tilde{\kappa}_d+ \left(
  \frac{r_K \left(\tilde{\kappa }_c \tilde{g}_d+\tilde{g}_c \tilde{\kappa
   }_d+N \tilde{\kappa }_c \tilde{\kappa }_d\right)+\frac{1}{2} N
   \left(\tilde{\kappa }_d^2-\tilde{\kappa }_c^2\right)-\tilde{g}_c
   \tilde{\kappa }_c+\tilde{g}_d \tilde{\kappa }_d}{r_K^2+1}+2
   \tilde{g}_d \tilde{\kappa }_d \right), \\
    \mu\partial_\mu \tilde{\kappa}_c &= -\epsilon \tilde{\kappa}_c+ \left(
    \tilde{\kappa }_c \tilde{g}_d+\tilde{g}_c \tilde{\kappa
   }_d+\frac{\tilde{\kappa }_c \tilde{g}_d+\tilde{g}_c \tilde{\kappa
   }_d+r_K \left(\frac{1}{2} N \left(\tilde{\kappa }_c^2-\tilde{\kappa
   }_d^2\right)+\tilde{g}_c \tilde{\kappa }_c-\tilde{g}_d \tilde{\kappa
   }_d\right)+N \tilde{\kappa }_c \tilde{\kappa }_d}{r_K^2+1}\right).
\end{align}

The one-loop fixed points are obtained by solving these flow equations. We have $\tilde\gamma=0$ at all fixed points. The equilibrium fixed points found from the $O(N) \times O(2)$ RG equations~\cite{Kawamura1998} solve our equations with $g_c^*/g_d^*=\kappa_c^*/\kappa_d^*=r_K$. These equilibrium fixed points all acquire an additional relevant direction, as discussed in the main text, and the fixed point that describes the phase transition at equilibrium becomes multicritical. This relevant direction is associated with the microscopic breaking of equilibrium conditions, which thus grows at large distances with a universal exponent. There are two new additional nonequilibrium fixed points toward which the flow is attracted, see Fig.~\ref{fig:FlowDiag}, and which therefore control the phase transition.
At these fixed points, the ratios $g_c^*/g_d^*$, $\kappa_c^*/\kappa_d^*$ and $r_K$ are not equal, and equilibrium conditions are violated. They exist for every $N>1$, which is not the case for the equilibrium ones~\cite{Kawamura1998}. 

However, there is no running of $r_K$ at one-loop order, and the values of the couplings at the fixed points all depend on the bare value $r_{K,B}$. There is therefore a (spurious) line of fixed points, which is a shortcoming of the one-loop equations. It is necessary to perform a two-loop analysis of the self-energies in order to obtain values of $r_K$ at the fixed points, and to fully characterize the latter even at first-order in $\epsilon$. This will also allow us to get the lowest-order expression of all critical exponents.

\subsubsection{b. Two-loop}

We now only consider the renormalization of $Z_t$, $Z_x$, $Z_D$ to get nontrivial renormalization of all parameters to lowest-order, which in turn allows us to get the running of $r_K$ via $\mu$-differentiation of~\eqref{eq:LinkCounter}. We thus only have to compute the contributions coming from the sunset diagrams, Fig.~\ref{fig:sunset1} and~\ref{fig:sunset2}. The first one contributes to $\Gamma^{(2)}_{\tilde\psi^* \psi}(p,\omega)$ a term
 \begin{align}
    \delta \Gamma_S(p,\omega)= -(\tilde g^2 \frac{N+1}{2}+ N\tilde\kappa^2+ 2 \tilde g\tilde\kappa )I_1(p,\omega) -\frac{1}{2}(\tilde g \tilde g^* \frac{N+1}{2} + \tilde\kappa\tilde\kappa^*+\tilde g\tilde\kappa^*+ \tilde g^*\tilde\kappa) I_2(p,\omega),
\end{align}
with
\begin{align}
    I_1(p,\omega) &= \int_{Q_1,Q_2} G^R(-Q_1-Q_2+P) G^K(Q_1) G^K(-Q_2) \\
    &= \int_{q_1,q_2}\frac{1}{q_1^2+\tilde\gamma }\frac{1}{q_2^2+\tilde\gamma }\frac{1}{- i \omega +3 \tilde\gamma+ (1+ i r_K)(q_2^2+(\boldsymbol{q}_1+\boldsymbol{q}_2-\boldsymbol{p})^2)+(1-i r_K)q_1^2}, \\
    I_2(p,\omega)&= \int_{Q_1,Q_2} G^R(Q_1+Q_2-P) G^K(Q_1) G^K(Q_2) \\
    &= \int_{q_1,q_2}\frac{1}{q_1^2+\tilde\gamma }\frac{1}{q_2^2+\tilde\gamma }\frac{1}{- i \omega +  3 \tilde\gamma+ (1+ i r_K)(q_1^2+q_2^2)+(1-i r_K)(\boldsymbol{q}_1+\boldsymbol{q}_2-\boldsymbol{p})^2},
\end{align}
where $Q_i=(\boldsymbol{q}_i,\omega_i)$ and $P=(\boldsymbol{p},\omega)$.
We only need the divergent parts at linear order in frequency and second order in $p$ that arise from the momentum integrations. They can be extracted using Feynman's parametrization and read as 
\begin{align}
    \partial_{-i\omega} I_1 &= -\frac{1}{(4\pi)^4 \epsilon}\frac{1}{(1+i r_K)^2}\ln (\frac{4}{3- i r_K}), \quad  \partial_{-i\omega} I_2 = -\frac{1}{(4\pi)^4 \epsilon}\frac{1}{(1-i r_K)^2}\ln (\frac{4}{(3- i r_K)(1+ i r_K)}),\\ 
    \partial_{p^2} I_1 &= -\frac{1}{(4\pi)^4 \epsilon}\frac{2-ir_K}{4(3-i r_K)} , \quad \partial_{p^2} I_2 = -\frac{1}{(4\pi)^4 \epsilon}\frac{1-ir_K}{6-2 i r_K}.
\end{align}
From 
$\partial_{p^2} \Gamma^{(2)}_{\tilde\psi^* \psi}(p,\omega) = 1+ \delta_{x} - \partial_{p^2} \left.\delta \Gamma_s\right|_{\textrm{sing.}}$ and $\partial_{-i\omega} \Gamma^{(2)}_{\tilde\psi^* \psi}(p,\omega) = 1+ \delta_{t} - \partial_{-i\omega} \left.\delta \Gamma_s\right|_{\textrm{sing.}}$, we obtain
\begin{align}
    Z_{x} &= 1- \frac{1}{\epsilon} \left[(\tilde g^2 \frac{N+1}{2}+ N\tilde\kappa^2+ 2 \tilde g\tilde\kappa )\frac{2-ir_K}{4(3-i r_K)} -\frac{1}{2}(\tilde g \tilde g^* \frac{N+1}{2} + \tilde\kappa\tilde\kappa^*+ \tilde g\tilde\kappa^*+ \tilde g^*\tilde\kappa)\frac{1-ir_K}{6-2 i r_K} \right],\\
    Z_t =& 1-\frac{\log \left(\frac{4}{(3-i r_K) (1+i r_K)}\right) \left(\tilde\kappa \tilde g^*+\tilde g
   \tilde\kappa ^*+\frac{1}{2} |\tilde g|^2 (N+1)+N |\tilde\kappa|^2 \right)}{2 (1-i
   r_K)^2 \epsilon }
   -\frac{\log \left(\frac{4}{3-i r_K}\right)
   \left(\frac{1}{2} \tilde g^2 (N+1)+2 \tilde g \tilde\kappa +\tilde\kappa ^2 N\right)}{(1+i
   r_K)^2 \epsilon }.
\end{align}

Similarly, evaluating the second sunset Fig.~\ref{fig:sunset2}, we find 

\begin{align}
\begin{split}
    \Gamma^{(2)}_{\tilde\psi_a^* \tilde\psi_a}(0,0)= & -2  (1 + \delta_D) + \left( \tilde g \tilde g^* \frac{(N+1)}{2}+\tilde\kappa  \tilde\kappa ^* N+\tilde g \tilde\kappa
   ^*+\tilde\kappa  g^*\right) \times \\ &\text{Re}\left(\int_{\boldsymbol{q}_1,\boldsymbol{q}_2}\frac{1}{q_1^2+\tilde\gamma }\frac{1}{q_2^2+\tilde\gamma }\frac{1}{(q_1+q_2)^2+\tilde\gamma }\frac{1}{3 \tilde\gamma+ (1+ i r_K)(q_1^2+ q_2^2) +(1-i r_K)(\boldsymbol{q}_1+\boldsymbol{q}_2)^2}\right).
   \end{split}
\end{align}
The poles of this integral can be found in~\cite{Risler2005} and in~\cite{Taeuber2014a} together with a detailed calculation (the function $L(r_K)$ which appears in (A16) of~\cite{Taeuber2014a} is obtained for $r_K>0$ and is in fact zero; its analytic continuation to negative $r_K$ is also zero). We finally obtain
 \begin{align}
    Z_D = 1-\frac{\left(\frac{1}{2} \tilde g \tilde g^* (N+1)+\tilde\kappa  \tilde\kappa ^* N+\tilde g \tilde\kappa
   ^*+\tilde\kappa  g^*\right)}{4
   \epsilon \left(1+r_K^2\right)} \left(3 \log
   \left(\frac{16}{(9+r_K^2)(1+r_K^2)}\right)+2 r_K \left(\arctan
   (r_K)+\arctan\left(\frac{r_K}{3}\right)\right)\right).
\end{align}
 
\begin{figure}[t]
    \centering
    \includegraphics{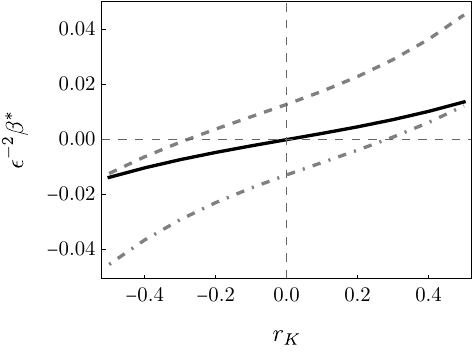}
    \caption{$\beta^*(r_K)=\beta_{r_K}(r_K, \tilde g^*(r_K),\tilde \kappa^*(r_K))$, plotted for different fixed points, i.e., different values of $\tilde g^*(r_K)$ and $\tilde \kappa^*(r_K))$  ($N=24$). The values of $r_K$ at the different fixed points are given by the zero of this function. The solid black line corresponds to any of the equilibrium fixed points up to a prefactor. Its zero is always found at $r_K^*=0$. The gray dashed and dashed-dotted lines correspond to the two nonequilibrium fixed points, for which $r_K^* \neq 0$. The fact that $r_K^*$ assumes opposite values at the two fixed points ensures that they are complex conjugates of each other.}
    \label{fig:beta_rk}
\end{figure}

We are now in position to get $Z_{r_K}$ using Eq.~\eqref{eq:LinkCounter},
from which we determine the $\beta-$function of $r_K$, $\beta_{r_K} = \partial_\mu r_K= -r_K \mu \partial_{\mu} Z_{r_K}/Z_{r_K}$,
\begin{equation}
\label{eq:Beta_rK}
\begin{split}
    \beta_{r_K} =&\frac{r_K}{2 \left(r_K^2+1\right)}\Big\{ \left(-\log \left(r_K^2+1\right)\right) \left(2 \left(\tilde{\kappa}_c^2+\tilde{\kappa}_d^2\right)+4 \tilde{g}_c \tilde{\kappa}_c+2 \tilde{g}_c^2+4 \tilde{g}_d \tilde{\kappa}_d+2
   \tilde{g}_d^2\right)\\ &+ \log \left(r_K^2+9\right) \Big(2 \tilde{g}_c \left(-6 \tilde{\kappa}_c r_K+2 \tilde{g}_d \left(r_K^2-1\right)-2 \tilde{\kappa}_d+2 \tilde{\kappa}_d
   r_K^2\right)+4 \tilde{g}_d \left(\tilde{\kappa}_c \left(r_K^2-1\right)+\tilde{\kappa}_d
   r_K\right)\\
   &+2 \left(-2 \tilde{\kappa}_c \tilde{\kappa}_d+2 \tilde{\kappa}_c \tilde{\kappa}_d
   r_K^2+r_K \left(\tilde{\kappa}_d^2-3 \tilde{\kappa}_c^2\right)\right)-6 \tilde{g}_c^2
   r_K+2 \tilde{g}_d^2 r_K\Big)
   \\ &-2 \log (4) \Big(2 \tilde{g}_c \left(-6 \tilde{\kappa}_c
   r_K+2 \tilde{g}_d \left(r_K^2-1\right)-2 \tilde{\kappa}_d+2 \tilde{\kappa}_d
   r_K^2\right) \\ 
   &+4 \tilde{g}_d \left(\tilde{\kappa}_c \left(r_K^2-1\right)+\tilde{\kappa}_d
   r_K\right)+2 \left(-2 \tilde{\kappa}_c \tilde{\kappa}_d+2 \tilde{\kappa}_c \tilde{\kappa}_d
   r_K^2+r_K \left(\tilde{\kappa}_d^2-3 \tilde{\kappa}_c^2\right)\right)-6 \tilde{g}_c^2
   r_K+2 \tilde{g}_d^2 r_K\Big) \\&+\frac{1}{r_K^2+9}\Big(\left(r_K^2+1\right) \big(-2 \tilde{g}_c
   \left(2 \tilde{\kappa}_c r_K+6 \tilde{g}_d+6 \tilde{\kappa}_d\right)+4 \tilde{g}_d \left(\tilde{\kappa}_d
   r_K \left(r_K^2+6\right)-3 \tilde{\kappa}_c\right) \\ 
   &+ 2 \left(-6 \tilde{\kappa}_c
   \tilde{\kappa}_d-\tilde{\kappa}_c^2 r_K+\tilde{\kappa}_d^2 r_K^3+6 \tilde{\kappa}_d^2
   r_K\right)-2 \tilde{g}_c^2 r_K+2 \tilde{g}_d^2 r_K
   \left(r_K^2+6\right)\big)\\& +\left(r_K^4+8 r_K^2-9\right) \arctan\left(r_K\right) \left(2 \left(\tilde{\kappa}_c^2+\tilde{\kappa}_d^2\right)+4
   \tilde{g}_c \tilde{\kappa}_c+2 \tilde{g}_c^2+4 \tilde{g}_d \tilde{\kappa}_d+2
   \tilde{g}_d^2\right)\\&+\left(r_K^2+9\right) \arctan
   \left(\frac{r_K}{3}\right) \big(4 \tilde{g}_c \left(\tilde{\kappa}_c
   \left(r_K^2-1\right)+4 \tilde{g}_d r_K+4 \tilde{\kappa}_d r_K\right)-4 \tilde{g}_d \left(-4
   \tilde{\kappa}_c r_K-3 \tilde{\kappa}_d+3 \tilde{\kappa}_d r_K^2\right)\\&+2 \left(8 \tilde{\kappa}_c \tilde{\kappa}_d r_K+\tilde{\kappa}_c^2 \left(r_K^2-1\right)-3 \tilde{\kappa}_d^2
   \left(r_K^2-1\right)\right)+2 \tilde{g}_c^2 \left(r_K^2-1\right)-6 \tilde{g}_d^2
   \left(r_K^2-1\right)\big)\Big)\Big\}.
\end{split}
\end{equation}

\begin{figure*}[t]
    \centering
\subfloat[][]{%
  \includegraphics{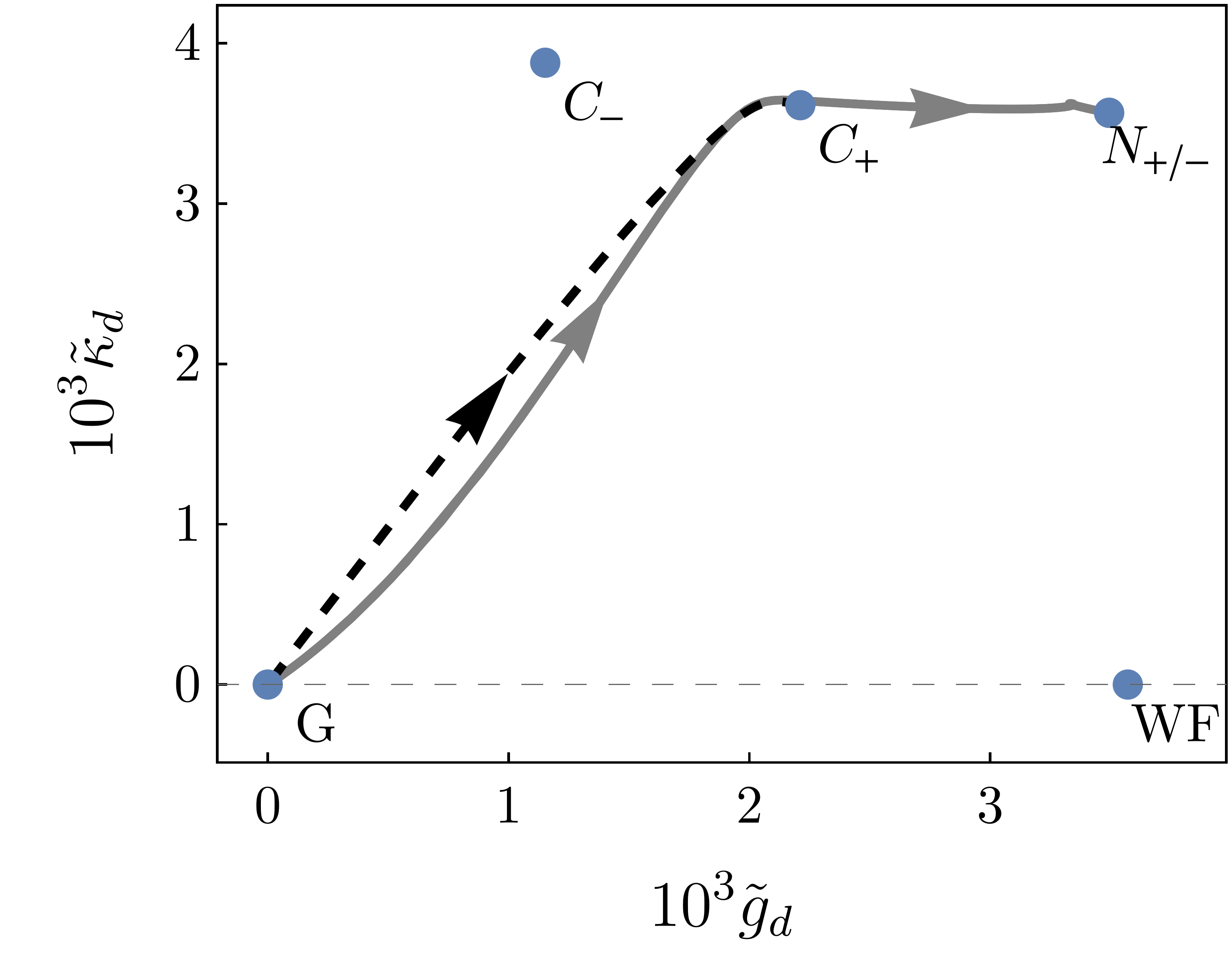}
}\hfil
\subfloat[][]{%
  \includegraphics{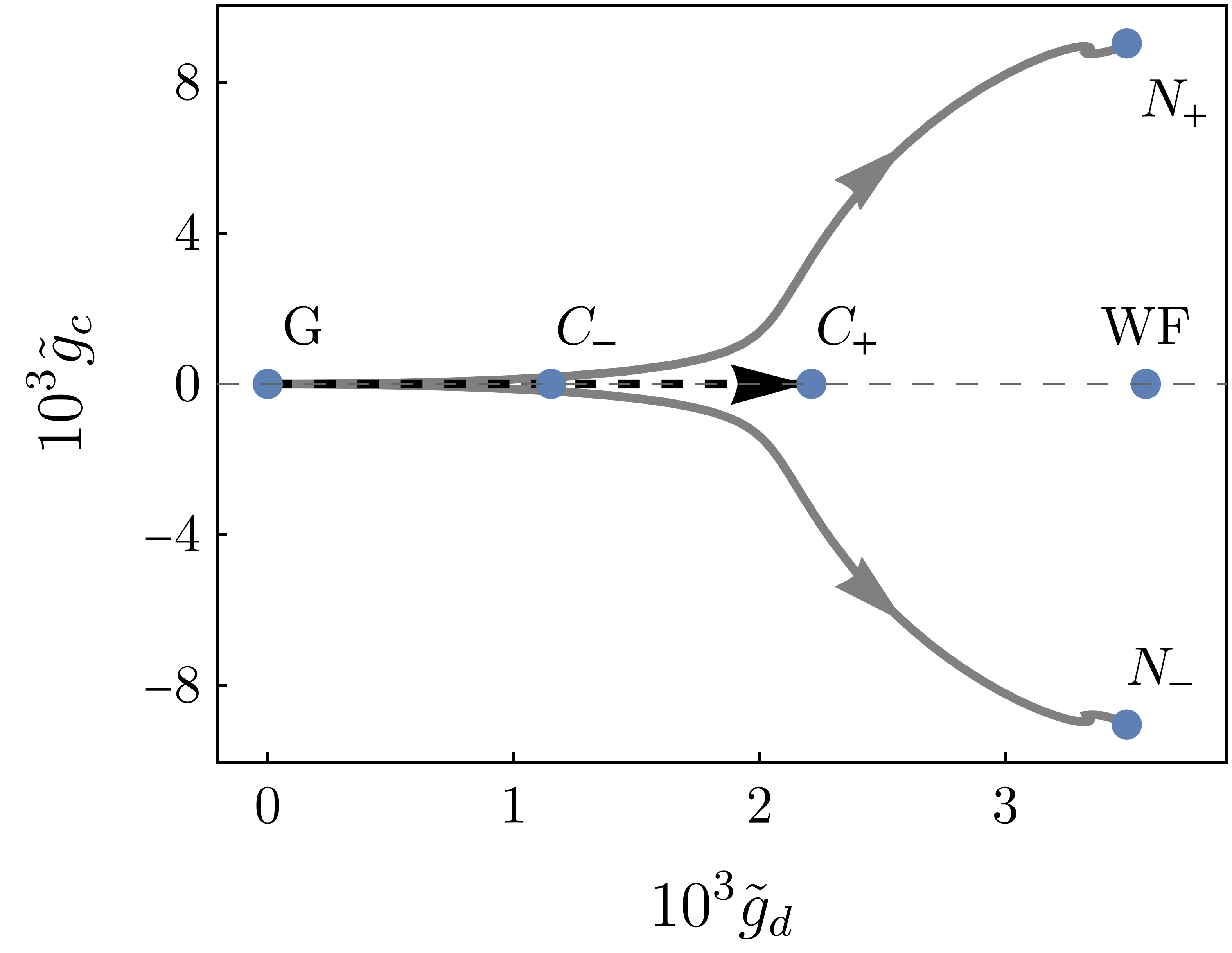}
}
    \caption{Flow diagram on the critical surface ($\tilde\gamma^*=0$) projected on the (a) $(\tilde g_d,\tilde \kappa_d)$ plane and (b) $(\tilde g_d,\tilde g_c)$ plane for $N=24$ and $\epsilon=0.1$. In addition to the Gaussian fixed point, there are three equilibrium fixed points (with $r_K^*= g_c^*= g_d^*=0$), labelled Wilson-Fisher (WF), chiral ($C_+$), and anti-chiral ($C_-$). The chiral one is attractive precisely at equilibrium, and the corresponding black dashed trajectory is attracted toward it. However, a small breaking of equilibrium conditions grows at larger scale, and the gray solid flow trajectories are attracted toward the new nonequilibrium fixed points $N_{\pm}$. They are complex conjugates of each other, see (b).}
    \label{fig:FlowDiag}
\end{figure*}

The full fixed points can now be obtained by using the following procedure: (i) The one-loop flow equations are solved as a function of $r_K$, e.g., $\tilde g^*(r_K)$. Note that these solutions are not fully fixed, since they still depend on $r_K$. (ii) The solutions can be directly fed into Eq.~\eqref{eq:Beta_rK} because only the one-loop corrections to interactions contribute to $\beta$ at two-loop order (i.e., at second order in $\epsilon$). (iii) The fixed point value $r_K^*$ is then obtained by solving $\beta^*(r_K)=\beta_{r_K}(r_K, \tilde g^*(r_K),\tilde \kappa^*(r_K))=0$. This can finally be injected back into the one-loop results to fully fix the solutions. We finally get the values of the coupling constants at the fixed point to order $O(\epsilon)$, e.g., $\tilde g^*(r_K^*)=O(\epsilon)$, and anomalous dimensions to order $O(\epsilon^2)$. We emphasize that we obtain $r_K^*= r_{K,2l}^*+ O(\epsilon)$ at two-loop order. 

For all equilibrium fixed points, i.e., for $\tilde g_c^*/\tilde g_d^*=\tilde\kappa_c^*/\tilde \kappa_d^*=r_K$, we find 
\begin{align}
    \beta^*(r_K)=& \frac{1}{2} \left(4\tilde g_d^* \tilde\kappa _d^*+(N+1)\tilde g_d^{*2}+2 N \tilde\kappa _d^{*2}\right) f(r_K),\\
   f(r_K)=&\left(r_K\log (\frac{16}{(r_K^2+1)(r_K^2+9)})+\left(r_K^2-1\right) \arctan
  \left(r_K\right)+\left(r_K^2+3\right) \arctan\left(\frac{r_K}{3}\right)\right),
\end{align}
which agrees with Eq.~(89) of~\cite{Taeuber2014a} when we set $N=1$.  Since the only zero of $f(r_K)$ is $r_K=0$ and $d\beta^*(r_K=0)/d r_K>0$, we find that $r_K^*=0$ for all equilibrium fixed points, no matter $N$. This reflects the fact that the equilibrium fixed points all have a purely dissipative dynamics. The fact that all imaginary parts are zero at the equilibrium fixed points, $g_c*=\kappa_c^*=r_K^*=0$, also means that these fixed points display an $O(N)\times O(2)$ symmetry, even if flow is initiated at equilibrium with $O(N)\times SO(2)$ symmetric initial conditions.

Away from equilibrium, the zero of $\beta^*(r_K)$ moves away of the origin, and $r_K$ reaches nonzero fixed-point values at the nonequilibrium fixed points, see Fig.~\ref{fig:beta_rk}. This implies that $\eta_c=\eta$, which means that coherent and dissipative parts have the same scaling. We get opposite nonzero values of $r_K$ for the two nonequilibrium fixed points that control the phase transition, ensuring that they are complex conjugates of each other. They thus describe mutually time-reversed coherent parts of the dynamics (e.g., $Z_c \to -Z_c$). The structure of the flow diagram is plotted in Fig~\ref{fig:FlowDiag}. Since the imaginary couplings do not vanish at these fixed points, they do not have an emergent $O(N)\times O(2)$ symmetry contrary to the equilibrium ones. They therefore describe a new universality class uniquely associated with the $O(N)\times SO(2)$ symmetry that controls the phase transition. The nonequilibrium fixed points have $\kappa_d>0$ for $N>N_{c} \sim 1.6+ O(\epsilon)$ and $\kappa_d<0$ otherwise. Since the Van der Pol phase exists only for $\kappa_d<0$, there is no fixed point associated with its onset. In perturbation theory, this is characteristic of a fluctuation-induced first-order transition~\cite{Amit2005}. The transition to the Van der Pol phase is therefore first-order for every $N>1$. Conversely, the transition to the rotating phase is second-order for all $N$, at least close to four dimensions, and described by the new universality class we find. 

Our calculation allows us to obtain all critical exponents to lowest order in $\epsilon$, see Tab.~\ref{tab:crit_exp} in the main text, and Tab.~\ref{tab:crit_exp_app}. The equilibrium exponents are obtained from analytical expressions, while the nonequilibrium ones are derived from numerical evaluation of the zeros of~\eqref{eq:Beta_rK}. In the equilibrium case where $g_c=g_d=r_K=0$, our equations reduce to the known two-loop $O(N)\times O(2)$ equations~\cite{Kawamura1988}, and the critical exponents agree as well. We obtain the $\eta_c$ exponents linked to the fade out of coherent dynamics, which was not known at equilibrium. 
The critical exponents of the new fixed points differ from the equilibrium case, as well as any other known universality class, confirming that we get a new universality class. We find that $\eta' \neq \eta$, which is only possible outside of equilibrium. Interestingly, $\eta'$ is complex and $r_K^*$ does not vanish. This gives rise to intriguing physical phenomena, such as \ac{RG} limit-cycle oscillations and absence of asymptotic decoherence~\cite{Marino2016a,Marino2016,Young2020}. This also implies a divergent temperature $T_\mu \sim \mu^{\eta-\textrm{Re}(\eta')}$, which can measured through correlation and response functions as shown above.

\begin{table*}
\renewcommand{\arraystretch}{1.25}
\setlength{\tabcolsep}{6pt}
\begin{tabular}{|c|c|c|c|c|c|c|}
    \hline
   $N$ & Phase & $\nu^{-1}-2$ &$\eta$& $z-2$ & $\eta'$ & $\eta_{c}$\\\hline
     $22$, eq. & Rot. & $-27/50\epsilon$ &  $ 0.0207 \epsilon^2$ & $0.0207 c\epsilon^2  $ & $\eta$ & $-0.0207 c' \epsilon^2 $\\\hline
     $22$, neq. & Rot. &  $-0.942\epsilon$ &$-0.142 \epsilon^ 2$ &$0.0055 \epsilon^ 2$  & $(0.00030+ 0.018i)\epsilon^2$ & $\eta$ \\\hline
     $3$, eq. & None & X & X&X & X &X\\\hline
     $3$, neq. & Rot. & $-1.27\epsilon$ & $-1.49 \epsilon^ 2$& $-0.017 \epsilon^ 2$&  $(-0.035+ 0.067i)\epsilon^2$   & $\eta$\\\hline
     $2$, eq. & vdP & $-\epsilon/2$ & $\epsilon^2/48$ & $c\epsilon^2/48$ &$\eta$  &$-c'\epsilon^2/48$\\\hline
     $2$, neq. & Rot. &  $-0.853\epsilon$ & $-0.353 \epsilon^ 2$ &$0.0072 \epsilon^ 2$ &  $(0.010+ 0.0070 i)\epsilon^2$& $\eta$\\\hline
\end{tabular}
\caption{Critical exponents for different values of $N$ in- and out-of-equilibrium to lowest nontrivial order in $\epsilon$. The equilibrium static results are reproduced, see~\cite{Kawamura1988}. The column ``Phase'' indicates the transition into which phase (rotating or Van der Pol (vdP)) is second order, while the other one is fluctuations induced first-order. For the $N=3$ equilibrium case, no attractive fixed point exists, and both phase transitions are first-order. We use $c=(6 \log(4/3)-1)$ and $c' =(4 \log(4/3)-1)$.}
\label{tab:crit_exp_app}
\end{table*}
\section{III. \texorpdfstring{$N=2$}{N=2} realizations}

Starting directly from the noisy Gross-Pitaevksii equation~\eqref{eq:EoM_complexApp}, one can identify additional realizations. Indeed, as its equilibrium counterpart, the $N=1$ version is known to give a Landau-type and widely applicable description of Bose-Einstein condensate with $U(1)$ symmetry in driven open dissipative systems~\cite{Sieberer2016}. In these systems, the drive pumps excitations, while the coupling with the environment opens the possibility for losses. In a semiclassical description, this two effects show up as noise and dissipation, the latter being associated with the imaginary part of the couplings. Since the number of excitation is not conserved, the dynamics is then given by a Gross-Pitaevksii equation of the form~\eqref{eq:EoM_complexApp} with additive noise.
When pumping exceeds loss, a bosonic mode condenses
at  a finite energy $\omega_0$, leading to a condensate that spontaneously breaks time-translation invariance, $\langle \Psi \rangle  \propto e^{-i \omega_0 t}$.

As far as the universal behavior is concerned, the correct effective field theory can thereby be identified by analyzing the symmetries of the system, and the construction is therefore robust. We thus generically expect condensation mechanisms in driven dissipative bosons with an $O(N)\times  U(1)\simeq O(N)\times SO(2)$ symmetry to be described by the new universality class discussed above.

A general scenario to get the noisy Gross-Pitaevskii equation~\eqref{eq:EoM_complexApp} with an $O(N=2)\times SO(2)$ symmetry is provided by driving two complex bosonic degrees of freedom $\psi_\pm$ connected by an exchange symmetry $\psi_+\leftrightarrow \psi_-$, as we now discuss in details in the light of two cases.

\subsection{1. Driven open exciton-polariton systems}
 
Exciton-polaritons are quasi-particles arising from light-matter interactions in a quantum-well placed inside a cavity. Using a laser field, the exciton-polaritons can be pumped into the system. They are also subject to loss processes such as photon leakage outside the cavity.
They have a polarization degree of freedom~\cite{Carusotto2013} that gives a two component complex bosonic field: $\boldsymbol{\psi}=(\psi_+,\psi_-)$. Depending on the experimental settings, there are situations where one only consider one of the polarization, and some cases where the two polarizations are used, see e.g.,~\cite{Ohadi2015} where a ferromagnetic transition is induced in polarization space. The former case corresponds to $N=1$ and has a $U(1)$ symmetry. 
The incoherent pumping and losses can be taken into account using a description in terms of a Lindblad equation~\cite{Carusotto2013,Sieberer2016}. This equation can then be recast as a path integral using the Keldysh formalism. The $U(1)$ case is discussed in details in \cite{Sieberer2016}. At large scale, the description becomes effectively semiclassical, and one recovers~\eqref{eq:EoM_complexApp} with $N=1$.
This approach can be directly generalized in the presence of the polarization degree of freedom.
In this case, the generic interacting Hamiltonian takes the form, for a contact interaction~\cite{Carusotto2013},
\begin{equation}
    \label{eq:H_ep}
    H_{\textrm{int}} = \int d^d \boldsymbol{x} \sum_{\sigma,\sigma'} \psi^*_\sigma(\boldsymbol{x})\psi_\sigma(\boldsymbol{x})V_{\sigma,\sigma'}\psi^*_{\sigma'}(\boldsymbol{x})\psi_{\sigma'}(\boldsymbol{x}), \quad V=\begin{pmatrix}
        V_t & V_s \\
        V_s & V_t
    \end{pmatrix},
\end{equation}
where $V_s$ and $V_t$ respectively describe scattering in the singlet and triplet polarization channel, and are generically different~\cite{Carusotto2013}. The procedure discussed above leads to an action of the form
\begin{equation}
    \label{eq:S_ep}
    S= \int_{\boldsymbol{x},t} \sum_\sigma \tilde\psi^*_{\sigma} (i\partial_t-\nabla^2 + \gamma_c+i\gamma_d)\psi_{\sigma}+\text{c.c}- 2 D \tilde\psi_\sigma \tilde\psi^ *_\sigma+ \sum_{\sigma,\sigma'} \psi^*_\sigma\psi_\sigma V_{\sigma,\sigma'}\psi^*_{\sigma'}\psi_{\sigma'},
\end{equation}
where all parameters (but $D$) now have an imaginary part arising from the drive and the dissipation, which respect the symmetry of the Hamiltonian. The single particle loss $\gamma_l$ and pumping $\gamma_p$ are related to the gap $\gamma_d=\gamma_l-\gamma_p$, and noise level $D=\gamma_l+\gamma_p$~\cite{Sieberer2016}. The transition is therefore reached when pumping exceed losses, i.e., when $\gamma_d<0$. 

Let us now analyze the symmetry class of this model. For $V_t \neq V_s$, there are two $U(1)$ symmetries, $U(1)_\pm$, and a $\mathbb{Z}_2$ symmetry: 
\begin{equation}
    \label{eq:sym_ep}
    U(1)_+ : \, \psi_+ \to \exp(i \theta_+)\psi_+, \quad  U(1)_- : \, \psi_- \to \exp(i \theta_-)\psi_-, \quad \mathbb{Z}_2: \,  \left\{\begin{array}{c}
          \psi_+ \to \psi_- \\ \psi_- \to \psi_+
    \end{array}\right., \quad \theta_\pm \in ]-\pi,\pi].
\end{equation}
One can alternatively write the $U(1)$ symmetries as
\begin{equation}
    \label{eq:sym_ep2}
    U(1)_s : \, \psi_\pm \to \exp(i \theta_s)\psi_\pm, \quad  U(1)_a : \, \psi_\pm \to \exp(\pm i \theta_a)\psi_\pm, \quad \mathbb{Z}_2: \,  \left\{\begin{array}{c}
          \psi_+ \to \psi_- \\ \psi_- \to \psi_+
    \end{array}\right., \quad \theta_\pm \in ]-\pi,\pi].
\end{equation}

We can now use the fact that $U(1)_s \simeq SO(2)$ and  $ U(1)_a \rtimes \mathbb{Z}_2 \simeq O(2)$. The semi-direct  product, $\rtimes$, reflects that the $Z_2$ and $U(1)_a$ transformations do not commute (while they commute with $U(1)_s$), exactly as rotations and reflections in $O(2)$. This additional $O(2)$ symmetry arises from the polarization degree of freedom.  The model therefore has $O(2)\times SO(2)$ symmetry. The driven Bose-Einstein condensation of polarized exciton polariton must therefore be in the universality class found in the main text for $N=2$. To make this fully explicit, we now show that Eq.~\eqref{eq:S_ep} can be mapped to Eq.~\eqref{eq:Action}. Guided by equation~\eqref{eq:sym_ep2}, we introduce symmetric and antisymmetric combinations of the fields~\cite{Pelissetto2002},
\begin{equation}
    \psi_+ \to \frac{1}{\sqrt{2}}(\psi_+ + i\psi_-),\quad \psi_- \to \frac{1}{\sqrt{2}}(\psi_- - i\psi_+),
\end{equation}
and the same relation for response fields.
This directly transforms~\eqref{eq:sym_ep} into~\eqref{eq:Action} with $g=V_s$, and $\kappa=(V_t-V_s)/2$ (or equivalently the Hamiltonian~\eqref{eq:H_ep} into~\eqref{eq:Ham_form}). These systems therefore realize our symmetry class for $N=2$ without any fine-tuning. Following this construction, driven-dissipative bosonic condensates with spins are good candidates to obtain $O(N)\times SO(2)$ symmetric complex Gross-Pitaevksii equations, much like spin one equilibrium bosons realize the $O(3) \times O(2)$ case, see e.g.,~\cite{Ohmi1998,Ho1998,Debelhoir2016}.

\subsection{2. Magnon condensation in YIG}

Yttrium iron garnet (YIG) is a ferromagnetic insulator with an exceptional long magnon lifetime. In  $\mu$m thick YIG films pumped by microwaves, a variant of the driven Bose-Einstein condensation of magnons has been observed \cite{Demokritov2006} at room temperature. Its mechanism is closely related to the one describing exciton-polariton systems, and has been described by complex Gross-Pitaevskii equations~\cite{Demokritov2006,Rezende2009}.
Due to the interplay of spin-orbit interactions and an in-plane magnetic field, the magnons in the system obtain a minimum in the band-structure at momenta $\pm k_0$ \cite{Rezende2009}. The two minima are equivalent by inversion symmetry. When  microwaves pump the system, the magnon density is increased, leading to a condensation at the momenta $\pm k_0$. The magnons can be represented via bosonic excitations~\cite{Rezende2009}, thus two complex fields $\boldsymbol{\psi}=(\psi_+,\psi_-)$ describe the system.

Both momentum (and energy) conservation~\cite{Rezende2009} and the rotating wave approximation restrict the possible interaction processes of the magnons in the infrared limit. Thus, the only allowed local interactions have the form 
\begin{align}
 \sum_{\sigma,\sigma'=\pm}  \int V_{\sigma\,\sigma'} \psi^\dagger_\sigma \psi_\sigma\psi^\dagger_{\sigma'} \psi_{\sigma'}
\end{align}
with $V_{++}=V_{--}$ and $V_{+-}=V_{-+}$ by inversion symmetry, which corresponds to the Hamiltonian~\eqref{eq:H_ep}. Because of the drive and decays of excited magnons, noise and dissipation are again generated. This leads then exactly to the field theory of Eq.~\eqref{eq:S_ep}. Therefore, this magnon condensation provides a realization of our effective field theory. 

Experiments of Nowik-Boltky {\it et al.} \cite{Nowik-Boltyk2012} show that 
the ordered state is inhomogeneous in space and therefore obtained from a superposition of the $\psi_+$ and $\psi_-$ condensate. We can use the mapping of Eq.~\eqref{eq:sym_ep2} to show that this state maps to the Van der Pol phase. 
Due to the finite gap $\omega_0$ and wavevector $k_0$, the condensate
$\langle \boldsymbol{\psi} \rangle$ oscillates in space and time, thus spontaneously breaking time-translational and space-translational invariances. 
These two symmetries have a $SO(2) \simeq U(1)$ structure in the presence of a finite scale, as discussed in the main text. This rationalizes the presence of the two $U(1)$ symmetries in the effective picture developed here.

\subsection{3. Experimental signatures}
Besides the direct {\em quantitative} comparison of critical exponents, two of our {\em qualitative} predictions are best suited to be compared to experiments. First, for the Van der Pol phase we predict a first-order transition, which can be detected by searching for hysteresis. Here, one can compare the behavior of the system when the pumping power is increased and reduced, respectively.
Second, the most significant qualitative prediction for the phase transition into the rotating phase is a diverging effective temperature, see Eq.~\eqref{eq:temp}. Experimentally, the effective temperature can be measured by comparing energy-gain and energy-loss of a system. In the language of Raman scattering, these processes are called Stokes and anti-Stokes lines. In equilibrium, their ratio is determined by detailed balance and given by $e^{-\hbar \omega/k_B T}$, where $\omega$ is the frequency which is probed. Out-of-equilibrium, one can use this relation as a way to define an effective temperature. Choosing for $\omega$ the rotation frequency of the rotating phase (or, equivalently, the energy of the condensing bosons), allows one to measure the effective temperature discussed in Eq.~\eqref{eq:temp}, which is predicted to diverge at criticality.

\end{document}